\documentclass[twocolumn]{aastex63}
\usepackage{natbib}
\usepackage{multirow}
\usepackage{textcomp}
\usepackage{gensymb}
\usepackage{lipsum}
\usepackage{blindtext}
\usepackage{mathtools}
\usepackage{tensor}
\usepackage{graphicx}

\emergencystretch=\maxdimen
\makeatletter

\shorttitle{Strong gravitational lensing effects by supermassive compact objects with regular spacetimes}
\shortauthors{J.~Kumar et al.}
\definecolor{MyDarkBlue}{rgb}{0,0.08,0.5}
\definecolor{MyDarkRed}{rgb}{0.7,0.02,0.02}
\definecolor{MyDarkGreen}{rgb}{0.0,0.7,0.0}

\usepackage{amsmath}

\begin{document}
\title{Testing strong gravitational lensing effects of  supermassive compact objects with regular spacetimes}
	%
	
	\correspondingauthor{Jitendra Kumar}
	\email{jitendra0158@gmail.com}
	\author[0000-0002-2724-9733]{Jitendra Kumar}
	\author[0000-0002-8539-5755]{Shafqat Ul Islam}
	\affiliation{ Centre for Theoretical Physics, Jamia Millia Islamia, New Delhi 110025, India}
	\author[0000-0002-0835-3690]{Sushant G. Ghosh}
	\affiliation{ Astrophysics Research Centre, School of Mathematics, Statistics and Computer Science, University of KwaZulu-Natal, Private Bag 54001, Durban 4000, South Africa}
		\affiliation{ Centre for Theoretical Physics, Jamia Millia Islamia, New Delhi 110025, India}

\begin{abstract}
We compare and contrast gravitational lensing, in the strong field limit, by the photon sphere in spherically symmetric regular electrically charged (REC)  black holes ($0<b\leq b_E$) and with those by corresponding REC no-horizon spacetimes ($b>b_E$).  Here, $b$ is an additional parameter due to the charge and the value $b=b_E \approx 0.226$ corresponds to an extremal black hole with degenerate horizons. Interestingly, the spacetime admits a photon sphere for $0<b\leq b_P \approx 0.247$ and an anti-photon sphere only for $b_E < b \leq b_P$. With no-horizon spacetime, images by lensing from the inside of the photon sphere ($u<u_{\text{ps}}$) can also appear. Interestingly,  for the case of  $u<u_{\text{ps}}$ the deflection angle $\alpha_D$ increases with $u$. We analyse the lensing observables by modeling compact objects Sgr A*, M87*, NGC 4649, and NGC 1332 as  black holes and no-horizon spacetimes. The angular position $\theta_{\infty}$ and photon sphere radius $x_{\text{ps}}$ decrease with increasing parameter $b$. Our findings suggest that the angular separations ($s$)  and magnification ($r$) of relativistic images inside the photon sphere may be higher than those outside. Moreover, the time delay for Sgr A* and M87* can reach $\sim$ 8.8809  and $\sim$ 12701.8 minutes, respectively, at $b = 0.2$, deviating from Schwarzschild black holes by $\sim$ 2.615  and  $\sim$ 4677 minutes. These deviations are insignificant for Sgr A* because it is too small, but they are sufficient for astronomical observation of M87* and some other black holes. With EHT bounds on the $\theta_{\text{sh}}$ of Sgr A* and M87*, within the $1 \sigma$ region, placing bounds on the parameter $b$,  our analysis concludes that the REC black holes agree with the EHT results in finite space, whereas the corresponding REC no-horizon spacetimes are completely ruled out.
\end{abstract}

\section{Introduction}
The weak cosmic censorship conjecture (CCC) by Penrose (\citeyear{Penrose:1969pc}), found that there can be no singularity visible from future null infinity, i.e., light rays originating from singularity are entirely blocked by the event horizon. Whereas the strong CCC prohibits its visibility by any observer. That means no light rays emanate out of the singularity, i.e., it is never naked, which mathematically precisely means spacetime should be globally hyperbolic. Despite almost half a decade of effort, we are still far from a general proof of CCC (for recent reviews and counterexamples, see \citep{Joshi:1987wg,Wald:1997wa,Singh:1997wa,Clarke_1994,Harada:1999jf,Joshi:2008zz}).  The gravitational lensing shows excellent potential for distinguishing black holes from
naked singularities based on qualitatively different lensing features
\citep{Virbhadra:2002ju,Virbhadra:1996cz,Virbhadra:1998kd,Virbhadra:2007kw} introducing the question as to  whether CCC could be tested observationally. Gravitational lensing is a meaningful astronomical means as it provides information about sources and lenses
and the large-scale geometry of the universe.

The idea of gravitational lensing was pioneered by Darwin (\citeyear{Darwin:1959}), who investigated photon trajectories passing in the environs of a black hole and showing the more significant deflection of a light ray, which in turn, guided an exact lens equation \citep{Frittelli:1999yf,Virbhadra:1999nm}. Later, Virbhadra {et al.} (\citeyear{Virbhadra:1998dy}), and Virbhadra and Ellis (\citeyear{Virbhadra:1999nm}) conducted a numerical investigation of lensing by static, spherically symmetric naked singularity and scrutinized lensing observables. Motivated by this, Perlick (\citeyear{Perlick:2003vg}) probed lensing in a spherically symmetric and static spacetime based on the light-like geodesic equation without approximations. It turns out that lensing by black holes is an excellent  astrophysical tool for investigating gravity's strong field features and providing information about the faraway dim stars.

Due to its importance, the gravitational lensing by various black holes has  attracted much attention in the past few decades with the functional, analytical techniques of Bozza (\citeyear{Bozza:2001xd}) based on the strong deflection of the light ray, showing that the deflection angle diverges logarithmically as light rays come to the photon sphere of a Schwarzschild black hole. The Bozza's recipe was applied to Reissner-Nordstr\"{o}m black holes \citep{Eiroa:2002mk}, and also to an arbitrary static, spherically symmetric metric \citep{Bozza:2002zj}. One of the motivations behind the diverse works on gravitational lensing is that the trajectories of light near black holes are related to the background geometry's essential features and properties, and also the anticipation that the relativistic images should test the gravity in the strong deflection limit led to applying this technique to diverse black hole metrics from  general relativity and modified theories \citep{Fernando:2002pa,Bozza:2002af,Majumdar:2004mz,Eiroa:2008ks,Bozza:2015dga,Sahu:2015dea,Islam:2021dyk}. Moreover, significant attention has been devoted to gravitational lensing by naked singularities or no-horizon spacetimes \citep{Gyulchev:2007cy}. The quantitative features of gravitational lensing can distinguish horizonless compact objects from black holes \citep{Gyulchev:2007cy,Gyulchev:2008ff}. Recently,
Shaikh {\it et al.} (\citeyear{Shaikh:2019itn}) performed an analytical investigation of strong gravitation lensing from such horizonless compact objects to obtain exact expressions of lensing observables for the images formed by lensing from the inside of the photon sphere and compared them with those formed from the outside.

This paper aims to investigate gravitational lensing by a class of spherical symmetric regular electrically charged (REC) black hole spacetimes \citep{Dymnikova:2004zc}, and probe how it differs from lensing by the corresponding no-horizon spacetimes. Assuming the supermassive black holes Sgr A* and M87* as the lens, we also compare REC black holes' observable signatures with those of the Schwarzschild black holes.  Interestingly, although it is feasible to detect some effects of the strong deflection lensing by the REC black holes with the Event Horizon Telescope (EHT) observations, it is tricky to distinguish two black holes as deviations are   $\mathcal{O}(\mu$as). We also use the EHT results from M87* and Sgr A* black hole shadow to constrain the deviation from the Kerr black hole and to assess the viability of REC regular black holes and no-horizon spacetimes.
By its very definition, the existence of a singularity means spacetime fails to exist, signaling a breakdown of the laws of physics.  In the absence of a well-developed definite quantum gravity permitting us to obtain the interior of the black hole and resolve it separately \citep{Wheeler:1964}, one must turn their attention to regular models, which are inspired by quantum assertions. The earliest idea by  Sakharov (\citeyear{Sakharov:1966}) and Gliner (\citeyear{Gliner:1966}) proposes that singularities could be avoided by matter with a de Sitter core, which could provide good discrimination at the final stage of gravitational collapse, replacing the future singularity \citep{Gliner:1966}. Based on this idea, Bardeen  (\citeyear{Bardeen:1968})  provided the first regular black hole solution \citep{Bardeen:1968} with horizons, but with no central singularity.  There has been a significant amount of  attention paid to the analysis and application of regular black holes \citep{Dymnikova:1992ux,AyonBeato:1998ub,Ayon-Beato:1999qin,Bronnikov:2000yz,Bronnikov:2000vy,Burinskii:2002pz,Dymnikova:2004zc,Hayward:2005gi,Bronnikov:2005gm,Hayward:2005gi,Zaslavskii:2009kp,Lemos:2011dq,Junior:2015fya,Fan:2016hvf,Bronnikov:2017tnz,Sajadi:2017glu,Toshmatov:2018cks}.   It has an additional parameter $g$ because of a magnetic monopole charge \citep{AyonBeato:1998ub}, apart from mass $M$ and encompasses the Schwarzschild metric as a particular case ($g=0$).
Thus, the black hole interior does not result in a singularity but develops a de Sitter-like region, eventually settling with a regular center: thus, its maximal extension is one of a Reissner\(-\)Nordstr$\mathrm{\ddot{o}}$m spacetime but with a regular center \citep{Borde:1994ai,Borde:1996df}. Also a siginificant  amount of  attention has been devoted to finding an extension, and uncovering of the  properties of Bardeen black holes has been reported \citep{Ansoldi:2008jw,Lemos:2011dq,Sharif:2011ja,Fernando:2012yw,Flachi:2012nv,Bambi:2014nta,Ulhoa:2013fca,Ghosh:2015pra,Schee:2015nua,Breton:2016mqh,Saleh:2018hba,Ali:2019myr,Dey:2018cws,Toshmatov:2019gxg,Kumar:2020bqf,Ghosh:2020tgy,Kumar:2020cve,Kumar:2020uyz,Islam:2022ybr}, and also in higher dimensional spacetimes \citep{Ali:2018boy,Kumar:2018vsm,Singh:2019wpu}.   Interestingly, the Bardeen model may also be interpreted as a solutions of Einstein equations with an electric source \citep{Rodrigues:2018bdc}.  Dymnikova (\citeyear{Dymnikova:2004zc}) proved the existence of electrically charged structures with a regular center, in which geometry, field, and stress-energy tensor are regular.

There is a good amount of motivation to constrain the electric charge of black holes, which has seen a multitude of efforts in this direction \citep{Takahashi:2005hy,EventHorizonTelescope:2021dqv,EventHorizonTelescope:2022xqj}. The charge of Sgr A* has also been constrained to be $Q\leq 3.1 \times 10^8$C using Chandra X-ray data \citep{Karouzos2018,Zajacek:2018ycb}. We intend to see whether we can improve this by using electromagnetic observations of supermassive black holes with the EHT results. The authors \citep{Kumar:2018ple,Ghosh:2020spb,KumarWalia:2022aop} demonstrated that the charged rotating regular black holes and corresponding no-horizon spacetimes agree with the EHT observations. Thus, charged rotating regular spacetimes and Kerr black holes are indiscernible in some parameter space, and one cannot rule out the possibility of the former being strong candidates for astrophysical black holes. A contribution to the characteristic lensing quantities because of the higher-order terms, viz., the gravitomagnetic terms considered in the lens potential, is analyzed. It turns out that  gravitomagnetic effects originate from the mass current in the lens and under some circumstances could give rise to results that could be detected with future observations \citep{Capozziello:2001yd, Capozziello:1999xz}.  It is because of the Rytov effect, i.e., the change in electromagnetic wave polarization in a smoothly inhomogeneous isotropic medium in a geometrical optics approximation \citep{rytov1938transition}. 

Therefore, it would be fascinating to explore the gravitation lensing by REC black holes/no-horizon spacetimes and compare it with a Schwarzschild black hole to place constraints on the deviation parameter.

This paper is organized as follows: In the Sec. \ref{sec2}, the spacetime structure of the REC black hole is briefly reviewed for completeness with a special focus on the energy conditions. The setup for gravitational lensing, including the lens equation, deflection angle, and the strong lensing coefficients, is given in Sec. \ref{sec3}. The strong lensing observables, i.e., the apparent size of the photon sphere (position of innermost image), the separation between the images and the difference in the brightness  between the relativistic images are also part of Sec. \ref{sec3}. Time delay for supermassive black holes Sgr A*, M87*, and those at the centres of 21 other galaxies have been estimated in Sec. \ref{sec4}.  Sec. \ref{sec5} is dedicated to the no-horizon spacetime wherein we discuss the lensing just inside the photon sphere and calculate the lensing observables in this case. The lensing observables for Sgr A*, M87*, NGC 4649, and NGC 1332 are discussed in Sec. \ref{sec6}. In Sec.~\ref{sec7}, we derive constraints on the rotating REC black hole parameter $b$ with the aid of EHT results from M87* and Sgr A* black hole shadow. We summarize our results and conclude the paper in Sec. \ref{sec8}.  

We have used units that fix the speed of light and
the gravitational constant $8 \pi G = c = 1$, but they are restored in the tables. 

\section{REC spacetimes}\label{sec2}

Here, we start with the action of gravity, minimally coupled to nonlinear electrodynamics, given by 
\begin{equation}\label{action}
S=\frac{1}{16\pi}\int{d^4 x\sqrt{-\mathbf{g}}({R-{\cal L}(F)}}); ~~ ~
F=F_{\mu\nu}F^{\mu\nu},                      
\end{equation}                                    
which contains Ricci scalar $R$, and
$F_{\mu\nu}=\partial_{\mu}A_{\nu}-\partial_{\nu}A_{\mu}$ is the electromagnetic field.
The gauge-invariant electromagnetic Lagrangian ${\cal L}(F)$ is an arbitrary function
of $F$, which in the weak field regime, should have the Maxwell limit, ${\cal L} \rightarrow F,~ {\cal
L}_F\rightarrow 1$.

Action (\ref{action}) gives the dynamic field equations
\begin{equation}\label{Lf}
 \nabla_{\mu}({\cal L}_F F^{\mu\nu});  ~~~ ~  \nabla_{\mu}(^{*}F^{\mu\nu})=0,                                     
\end{equation}
where ${\cal L}_F=d {\cal L}/dF$. In the spherically symmetric
case, the only nonzero components of $F_{\mu\nu}$ are a radial
electric field $F_{01}=-F_{10}=E(r)$ and a radial magnetic field
$F_{23}=-F_{32}$.
The Einstein equations take the form \citep{Bronnikov:2000vy}
\begin{equation}\label{EE}
G^{\mu}_{\nu}=-T^{\mu}_{\nu}=2 {\cal L}_F
F_{\nu\alpha}F^{\mu\alpha}-\frac{1}{2}\delta_{\nu}^{\mu} {\cal L}.
\end{equation}                                                 
The density and pressures for electrically charged source are given by
\begin{equation}\label{emt}
\rho=-p_r=\frac{1}{2}{\cal L}-F{\cal L}_F;~~~
p_{\perp}=-\frac{1}{2}{\cal L},
\end{equation}
and the scalar curvature is $R=2({\cal L}-F{\cal L}_F)=2(\rho - p_{\perp})$.
We consider the general spherically symmetric metric ansatz
 \begin{equation}\label{metric}
 ds^2=A(r) dt^2 - \frac{dr^2}{B(r)} - r^2 d\Omega^2,
\end{equation}                                              such that
\begin{equation}\label{Fr}
A(r)=\frac{1}{B(r)}=1-\frac{2{\cal M}(r)}{r}; ~~~
 {\cal
M}(r)=\frac{1}{2}\int_0^r{\rho(x)x^2dx},
\end{equation}
where $d\Omega^2$ is the line element on a unit two-sphere.
The solution we are interested in can be derived by choosing the density and pressure as
\begin{equation}\label{Esol}
\rho(r)=\frac{q^2}{(r^2+g^2)^2};~~~~
p_{\perp}=\frac{q^2 (r^2-g^2)}{(r^2+g^2)^3}; ~~~ g= \frac{\pi q^2}{8m},
\end{equation}
where $g$ is the electromagnetic radius of the nonlinear electrodynamics \citep{Dymnikova:2004zc}. 
Then the Lagrangian and its derivative are
\begin{equation}\label{Lsol}
{\cal L}=\frac{2q^2(g^2-r^2)}{(r^2+g^2)^3}; ~~~ {\cal
	L}_F=\frac{(r^2+g^2)^3}{r^6}.
\end{equation}
Integration of Equation (\ref{EE}) with the density Equation (\ref{Esol}) leads to the solution 
\begin{equation}\label{solF}
A(r)=1/B(r)=1-\frac{4m}{\pi r}\biggl[\arctan \left(\frac{r}{g}\right)-\frac{r
	g}{r^2+g^2}\biggr],
\end{equation}
where $m$ is the mass of the black hole and $g$ is a parameter that appears due to the coupling of nonlinear electrodynamics to general relativity \citep{Dymnikova:2004zc}. The metric Equation (\ref{solF}) is spherically symmetric and is also asymptotically flat as $\displaystyle{\lim_{r \to \infty}}$ $ A(r)=B(r)=1$ and though regular everywhere for nonzero $g$, reduces to the singular Schwarzschild metric  in the absence of nonlinear electrodynamics ($g = 0$). Asymptotically for $r \gg g$, metric Equation (\ref{solF}) reads as
\begin{equation}\label{solL}
A(r)=1-\frac{2m}{r}+\frac{q^2}{r^2}-\frac{2}{3}\frac{q^2
	g^2}{r^4},
\end{equation}
and has a Reissner-Nordstr\"om limit as $r\rightarrow \infty$. At small values of $r$, $r \ll g$ we get de Sitter asymptotic  
with the cosmological constant
$\Lambda=q^2/g^4$
which gives a useful expression for a cutoff on self-energy density
by the finite value of the electromagnetic density $T_0^0\;
 (r\rightarrow 0)$ associated with the cosmological constant $\Lambda
= 8\pi \rho(0)$, which emerges at the regular center. The mass of electromagnetic origin is related to this cutoff by
$m=\pi^2\; \rho(0)\; g^3$.

We approach the regularity problem of the solution by studying the behavior of invariant $R= R_{a}\,R^{a}$,  Ricci scalar  $\mathcal{R}=R_{ab}\,R^{ab}$, ($R_{ab}$ Ricci tensor), and the Kretschmann scalar $K=R_{abcd}\,R^{abcd}$($R_{abcd}$ Reimann tensor). They take the form
\begin{equation}
    R~= \frac{32mg^3}{\pi \left(r^2+g^2\right)^3},  ~~~~\\
\mathcal{R} = \frac{256m^2g^2\left(r^4+g^4\right)}{\pi^2 \left(r^2+g^2\right)^6}~,
\end{equation}
\begin{equation*}
  K= \frac{192m^2}{\pi^2 r^6 \left(r^2+g^2\right)^6} \left(K_1 + K_2 \right),
    \end{equation*}
where \\   
$K_1= \left(r^2+g^2\right)^6\left[\arctan\left(\frac{r}{g}\right)\right]^2 \\ -2\left[\arctan\left(\frac{r}{g}\right)\right] rg\left(r^2+g^2\right)^3\left(3r^4+\frac{8}{3}r^2g^2+g^4\right)$, \\
$K_2= r^2g^2\left(\frac{5}{3}r^4+\frac{4}{3}r^2g^2+g^4\right) \left(7r^4+4r^2g^2+g^4\right)$, 

which are obviously well-behaved everywhere including as $r\to 0$, when $g, m\neq0$ and $r\to 0$. At the center, they vanish. Thus, the spacetime Equation (\ref{metric}) is regular or nonsingular. 

\begin{figure} 
		   \includegraphics[scale=0.8]{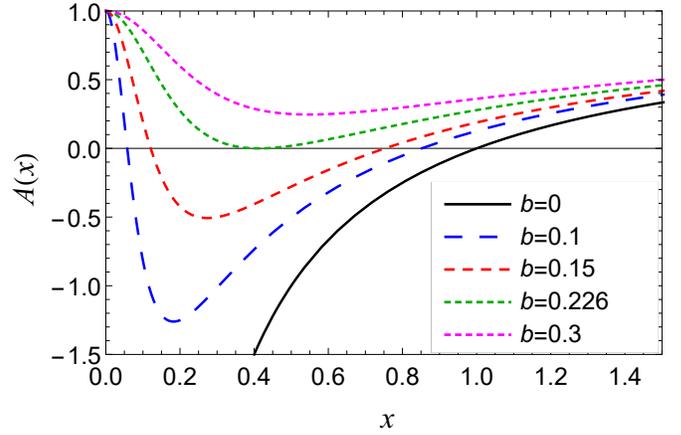}
	\caption{Plot showing $A(x)$ vs. $x$; for $0<b\leq b_E$ one has a black hole with two horizons while $b>b_E$ corresponds to no-horizon spacetime. Case $b=0$ corresponds to the Schwarzschild black hole. }\label{fig1}
\end{figure}

\begin{figure} 
		   \includegraphics[scale=0.8]{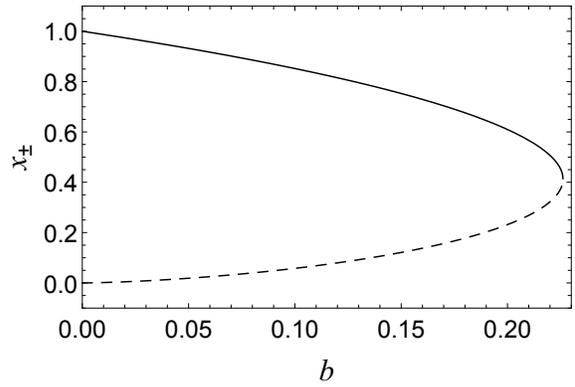}
	\caption{Locations of event horizon $x_{+}$ (solid line) and Cauchy horizon $x_{-}$ (dashed line) for REC black holes.}\label{fig2}
\end{figure}

\subsection{Energy conditions and regularity} 
Next, we check the status of the various energy conditions using the prescription of Hawking \& Ellis (\citeyear{Hawking:1973uf}) \citep{Kothawala:2004fy,Ghosh:2008zza,Hawking:1973uf}. 
The Einstein equations governing the stress-energy tensor $T_{\mu\nu}$ are given by the Equation (\ref{EE}), which leads to
\begin{eqnarray}\label{pr}
\rho = \frac{8mg}{\pi \left(r^2+g^2\right)^2} = -P_{r},\\
P_{\theta} = P_{\phi}= \frac{8mg\left(r^2-g^2\right)}{\pi \left(r^2+g^2\right)^3} . 
\end{eqnarray}

The weak energy condition (WEC), $T_{\mu\nu}\xi^{\mu}\xi^{\nu}\geq 0$ for any timelike vector
$\xi^{\mu}$, which is equivalent to the findings in Hawking \& Ellis (\citeyear{Hawking:1973uf}).
\begin{equation}\label{1a}
\rho\geq 0;  ~~ ~\rho + P_i \geq 0,~~~ (i=r,\; \theta,\; \phi)
\end{equation}

guarantees that the energy density as measured by any local
observer is non-negative.  Using the Eq.~(\ref{solF})
\begin{equation}\label{1aa}
    \rho+P_{\theta}= \rho+P_{\phi} = \frac{16mgr^2}{\pi \left(r^2+g^2\right)^3}.
\end{equation}
Hence, $\rho\geq 0$ and  Eq.(\ref{1aa}) imply that the WEC is satisfied everywhere. 

\begin{figure*} 
	\begin{centering}
		\begin{tabular}{c c}
     \includegraphics[scale=0.8]{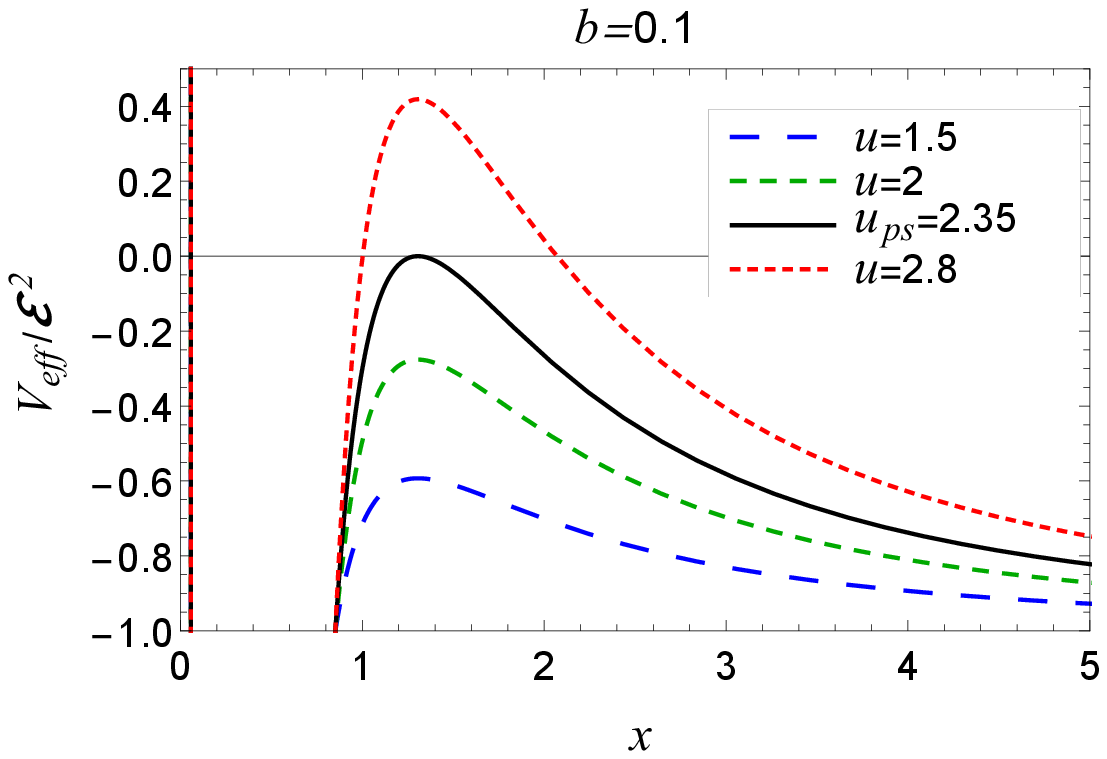}	\hspace{0.8cm}
	\includegraphics[scale=0.7]{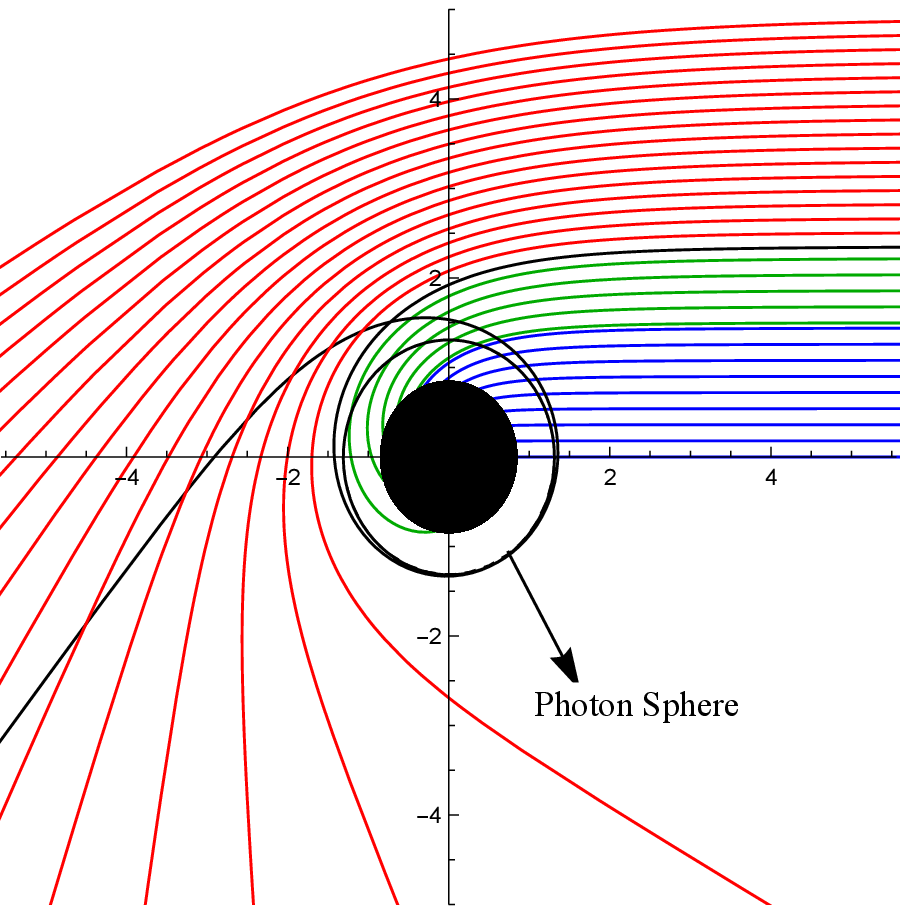}
			\end{tabular}
	\end{centering}
	\caption{(a) (left) Variation of the effective potential $V_{\text{eff}}$ for REC black holes as a function of  radial coordinate $x$. The photons with critical impact parameter ($u_{\text{ps}}$) (black solid curve) make unstable circular orbits. (b) (right) Trajectories of photons for REC black holes with different impact parameters $u$. Photons having impact parameter $u>u_{\text{ps}}$ (red curves) are scattered to infinity, while photons impacted by parameter $u<u_{\text{ps}}$ (green and blue curves) are absorbed in the black hole. Photons with an impact parameter almost close to the critical impact parameter $u \approx u_{\text{ps}}$ (black curve) make several loops around the black hole before being scattered to infinity. }\label{fig3}	
\end{figure*}
\begin{figure*} 
	\begin{centering}
		\begin{tabular}{c c}
     \includegraphics[scale=0.8]{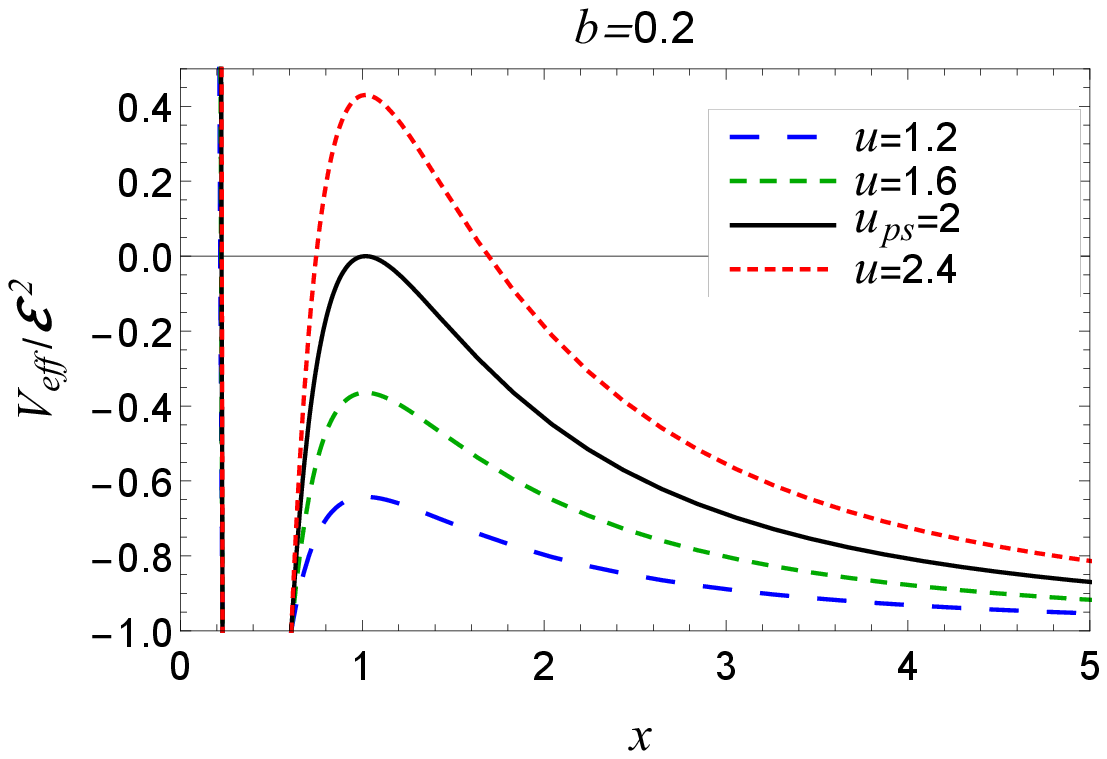}	\hspace{0.7cm}
	\includegraphics[scale=0.7]{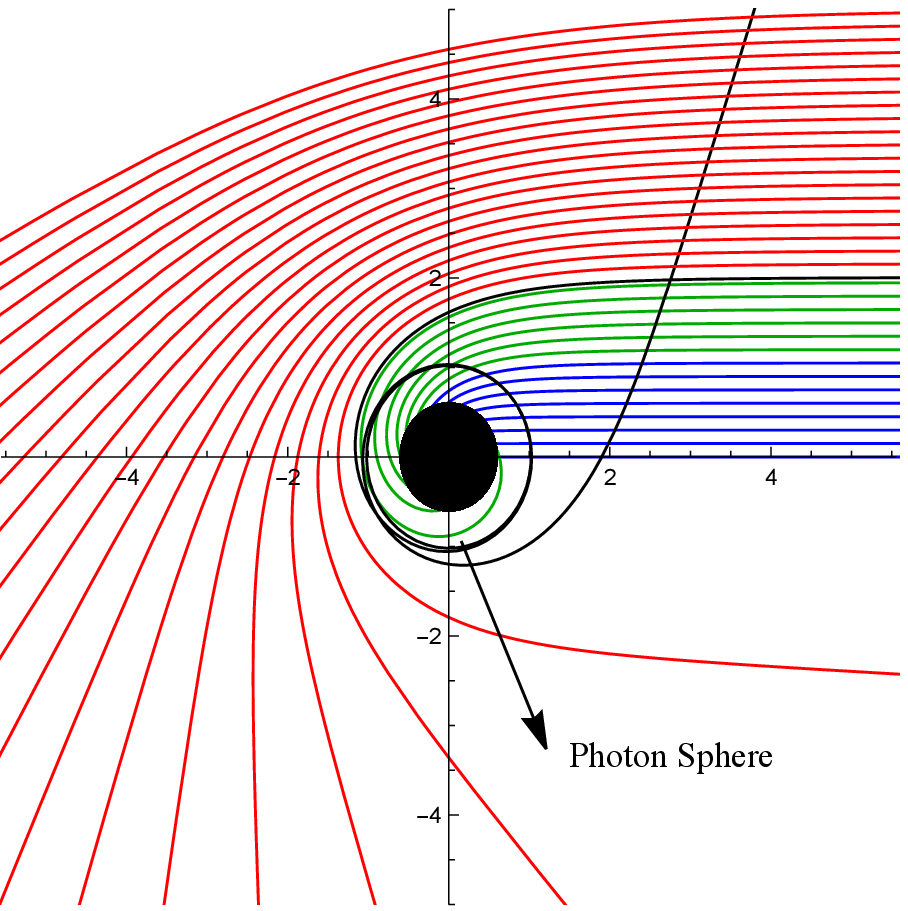}
			\end{tabular}
	\end{centering}
	\caption{(a) (left) Variation of the effective potential $V_{\text{eff}}$ for REC black holes as a function of  radial coordinate $x$. The photons with critical impact parameter ($u_{\text{ps}}$) (black solid curve) make unstable circular orbits. (b) (right) Trajectories of photons for REC black holes having different impact parameters $u$. A Photon with impact parameter $u>u_{\text{ps}}$ (red curves) is scattered to infinity, while the photon impacted by parameter $u<u_{\text{ps}}$ (green and blue curves) is absorbed in the black hole. A photon with an impact parameter almost close to the critical impact parameter $u \approx u_{\text{ps}}$ (black curve) makes several loops around the black hole before being scattered to infinity.}\label{fig3a}	
\end{figure*}
Next, the dominant energy condition  (DEC), $T^{00}\geq|T^{ki}|$ for
each $i=r,\; \theta,\; \phi$, which holds if and only if \citep{Hawking:1973uf}
\begin{equation}\label{1b}
\rho\geq0;~~\rho + P_i \geq 0;~~\rho-P_i \geq 0,~~~(i=r,\; \theta,\; \phi)
\end{equation}
includes WEC and requires that each principal pressure $P_k=-T^i_i$
never exceeds the energy density. To check the DEC, we first calculate
\begin{equation}\label{1bb}
\rho - P_{\theta} = \rho - P_{\phi} = \frac{16mg^3}{\pi \left(r^2+g^2\right)^3}.
\end{equation}
Thus, the DEC holds because of the WEC and   Eq.(\ref{1bb}). 
Lastly, the  strong energy condition (SEC)  requires \citep{Hawking:1973uf}
\begin{equation}\label{1c}
\rho + \sum_i{P_i} \geq 0
\end{equation}
and defines the sign of the gravitational acceleration. Thus, the SEC  requires $\rho+P_{r}+2~P_{\theta}\geq 0.$
We find that 
\begin{equation}
    \begin{aligned}\label{st}
    \rho+P_{r}+2~P_{\theta} = \frac{16mg\left(r^2-g^2\right)}{\pi \left(r^2+g^2\right)^3}
\end{aligned}
\end{equation} 
which means that the SEC is satisfied when $r^2 > g^2.$

\section{Strong Gravitational Lensing in REC Black Hole Spacetime}\label{sec3}
We begin with rewriting the metric Equation (\ref{metric}) by redefining the quantities $r$, $q$ and $t$ in the units of radius $2m$ as  
\begin{eqnarray}\label{metrcomp}
&& A(x)=\frac{1}{B(x)}=1 -\frac{2}{\pi x}\left[\arctan \left(\frac{x}{b}\right)- \frac{x b}{x^2+b^2} \right],\;\;\;\;\nonumber \\ && C(x)=x^2,
\end{eqnarray}
where $b=g/2m$, $x=r/2m$ and $t=t/2m$.
An elementary analysis of $A(x)=0$ suggests a critical value of $b_E \approx 0.226$ corresponding to an extremal REC black hole (see Figure \ref{fig1}). On the other hand, $A(x)=0$ has no positive root for $b>b_E$ and admits two distinct positive roots for $b<b_E$. They, respectively, describe no-horizon REC spacetime and non-extremal black holes. The critical point $b=b_E=0.226$ corresponds to an extremal black hole with $A(x)=0$ having equal roots (see Figures \ref{fig1} and \ref{fig2}). Further, in no-horizon REC spacetime, the photon sphere can exist when $b_E < b \leq b_P \approx 0.247$ and no photon sphere exists when $b > b_P$.

\begin{figure*}[t] 
	\begin{centering}
		\begin{tabular}{c c}
		    \includegraphics[scale=0.85]{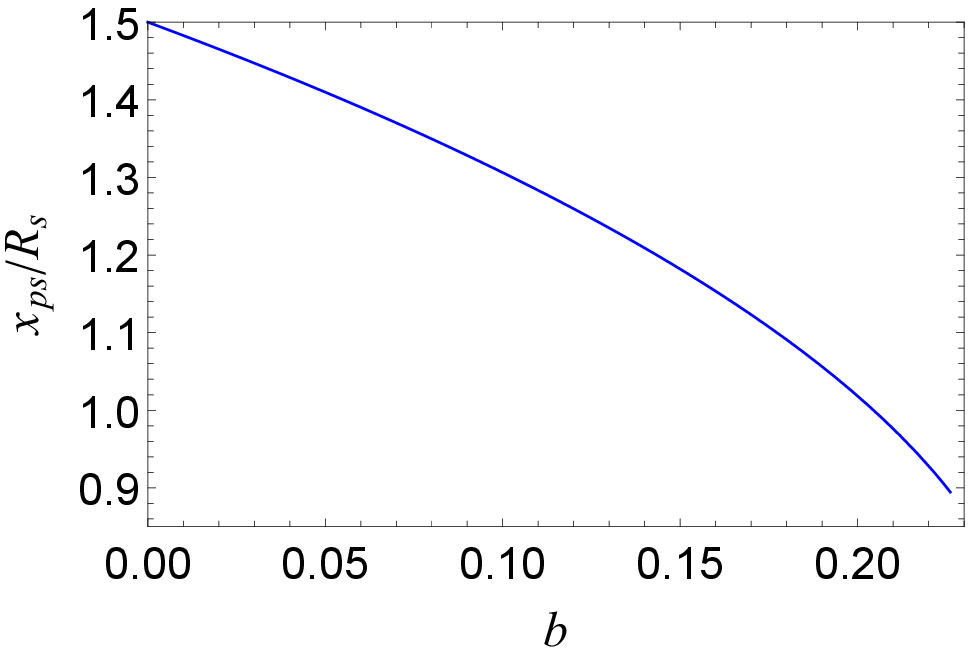}\hspace{0.5cm}
		    \includegraphics[scale=0.85]{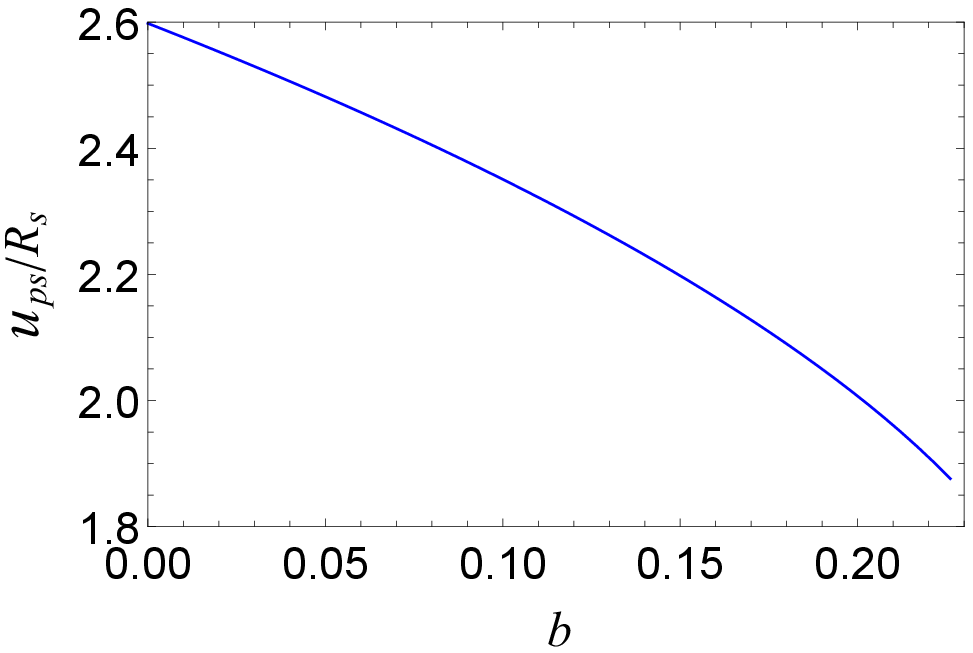}
		    \end{tabular}
	\end{centering}
	\caption{Behavior of the photon sphere radius $x_{\text{ps}}$ (left) and the critical impact parameter $u_{\text{ps}}$ (right) for REC black hole spacetime as a function of the parameter $b$. As $b \to 0$, the values  $x_{\text{ps}} \to 1.5$ and $u_{\text{ps}} \to 2.598$ correspond to a Schwarzschild black hole \citep{Bozza:2002zj}. }\label{fig4}
\end{figure*}

The strong field gravitational lensing is governed by the deflection angle and a lens equation. The null geodesic equation reads as

\begin{equation}\label{Veff}
\left(\frac{dx}{d\tau}\right)^2\equiv \dot{x}^2={\cal E}^2-\frac{\mathcal{L}^2A(x)}{C(x)},
\end{equation}
where the energy $\mathcal{E}=-p_{\mu}\xi^{\mu}_{(t)}$ and angular momentum $\mathcal{L}=p_{\mu}\xi^{\mu}_{(\phi)}$, with $\xi^{\mu}_{(t)}$ and $\xi^{\mu}_{(\phi)}$ are, respectively, the Killing vectors due to time-translational and rotational invariance. The radius of the photon sphere is the largest root found by Virbhadra \& Ellis (\citeyear{Virbhadra:1999nm}) and Bozza (\citeyear{Bozza:2002zj}):
\begin{equation}
\frac{C'(x)}{C(x)}=\frac{A'(x)}{A(x)},\label{root} 
\end{equation}
which is depicted in the Figure~\ref{fig4}. At the closest approach distance $x_0$, $\dot{x}|_{x=x_0}=0$ or $V_{\text{eff}}(x_0)=0$, which leads to 
\begin{equation}\label{impact}
u\equiv \frac{\cal L}{\cal E} =\sqrt{\frac{C(x_0)}{A(x_0)}}.
\end{equation} 
The effective potential from the radial Equation (\ref{Veff}) reads as
\begin{equation}\label{veff}
\frac{V_{\text{eff}}}{{\cal E}^2} =  \frac{u^2}{x^2}\left[1 -\frac{2}{\pi x}\left(\arctan\left(\frac{x}{b}\right)- \frac{x b}{x^2+b^2} \right)\right]-1.    
\end{equation}
It turns out that a light ray exists in the region where 
$V_{\text{eff}} \leqslant 0$ (see Figures \ref{fig3} and \ref{fig3a}). Further, one can define an unstable (or a stable circular orbit) satisfying  $V_{\text{eff}}=V'_{\text{eff}}(x)=0$ and  
$V''_{\text{eff}}(x_{\text{ps}}) < 0$ (or $V''_{\text{eff}}(x_{\text{ps}}) > 0$). 
For REC spacetime, we find that $V''_{\text{eff}}(x_{\text{ps}}) < 0$, which corresponds to the unstable photon circular orbits (see Figures \ref{fig3} and \ref{fig3a}).

\begin{figure*}[t] 
	\begin{centering}
		\begin{tabular}{cc}
		    \includegraphics[scale=0.85]{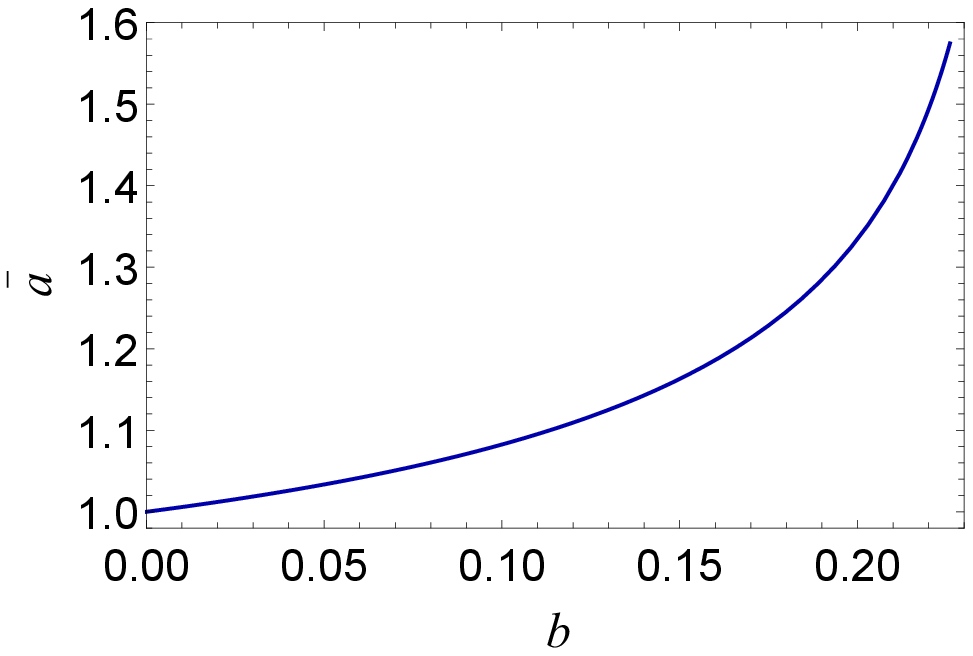}\hspace{0.5cm}
		    \includegraphics[scale=0.85]{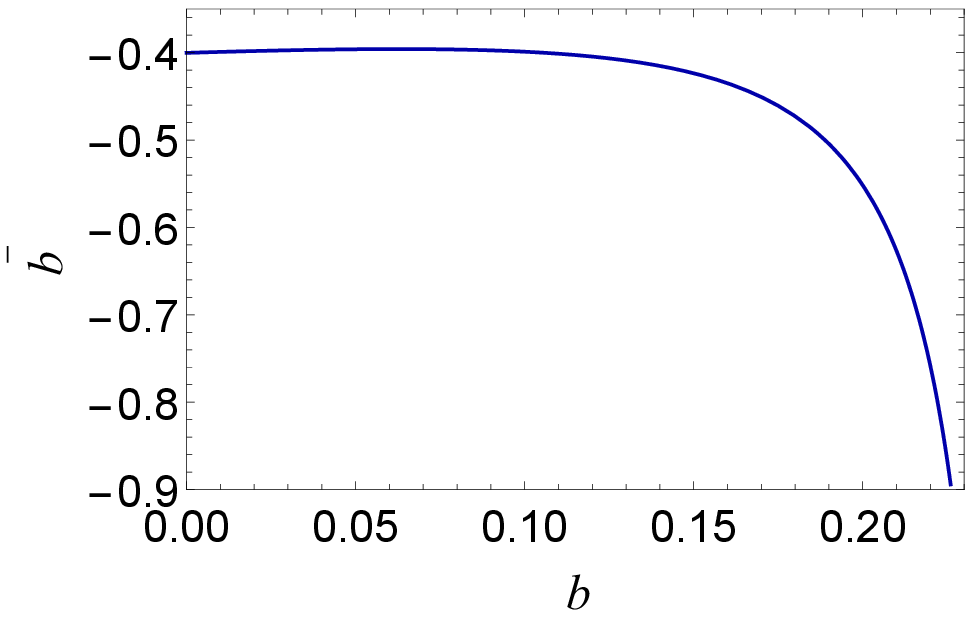}
			\end{tabular}
	\end{centering}
	\caption{Behavior of strong lensing coefficients $\bar{a}$ and $\bar{b}$ for REC black hole spacetime as a function of the parameter $b$. $\bar{a}=1$ and $\bar{b}=-0.4002$ at $b=0$ correspond to the values of a Schwarzschild black hole.}\label{fig5}	
\end{figure*}
The deflection angle becomes unboundedly large at $x_0=x_{\text{ps}}$ and is finite only for $x_0>x_{\text{ps}}$.
The critical impact parameter $u_{\text{ps}}$ is defined as \citep{Bozza:2002zj}
\begin{equation}\label{crimp}
u_{\text{ps}}=\sqrt{\frac{C(x_{\text{ps}})}{A(x_{\text{ps}})}},
\end{equation} 
and depicted in Figure \ref{fig4}. This is the impact parameter for which the photon approaches the black hole with the closest approach distance $x_0=x_{\text{ps}}$ and revolves around it in an unstable circular orbit of radius $x_{\text{ps}}$. The photons that have impact parameter $u < u_{\text{ps}}$ fall into the black hole, while photons with impact parameter $u > u_{\text{ps}}$ reach the minimum distance $x_0$ near the black hole and then are symmetrically scattered to infinity.

\begin{figure*}[t] 
	\begin{centering}
		\begin{tabular}{cc}
		   \includegraphics[scale=0.77]{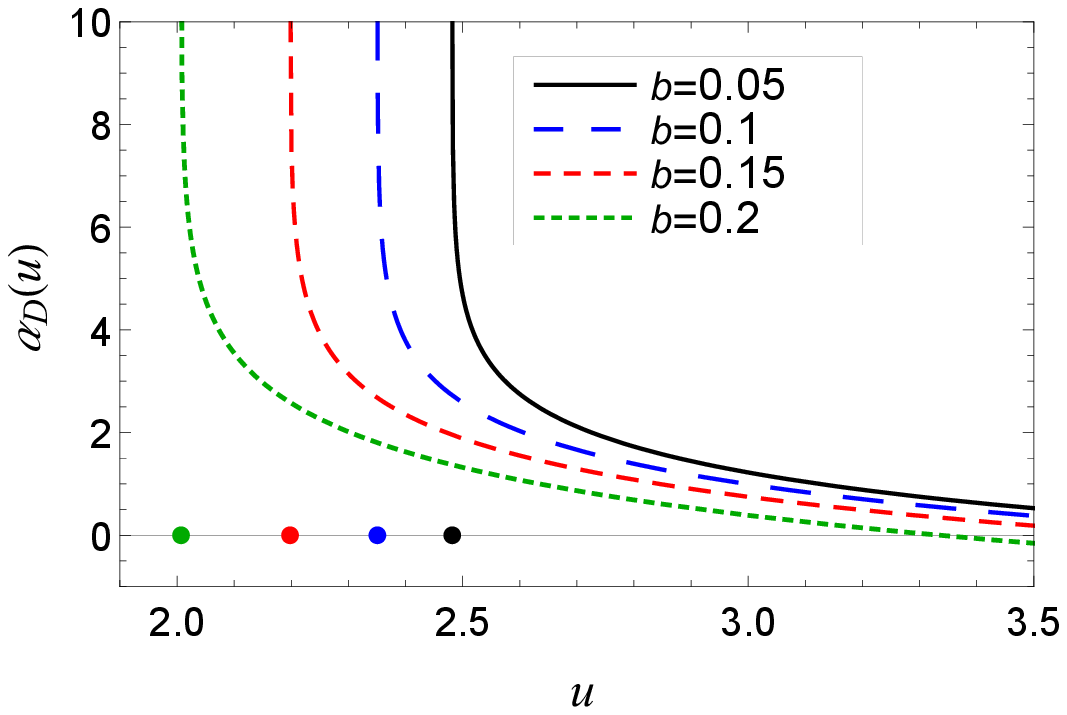}\hspace{0.5cm}
		   \includegraphics[scale=0.695]{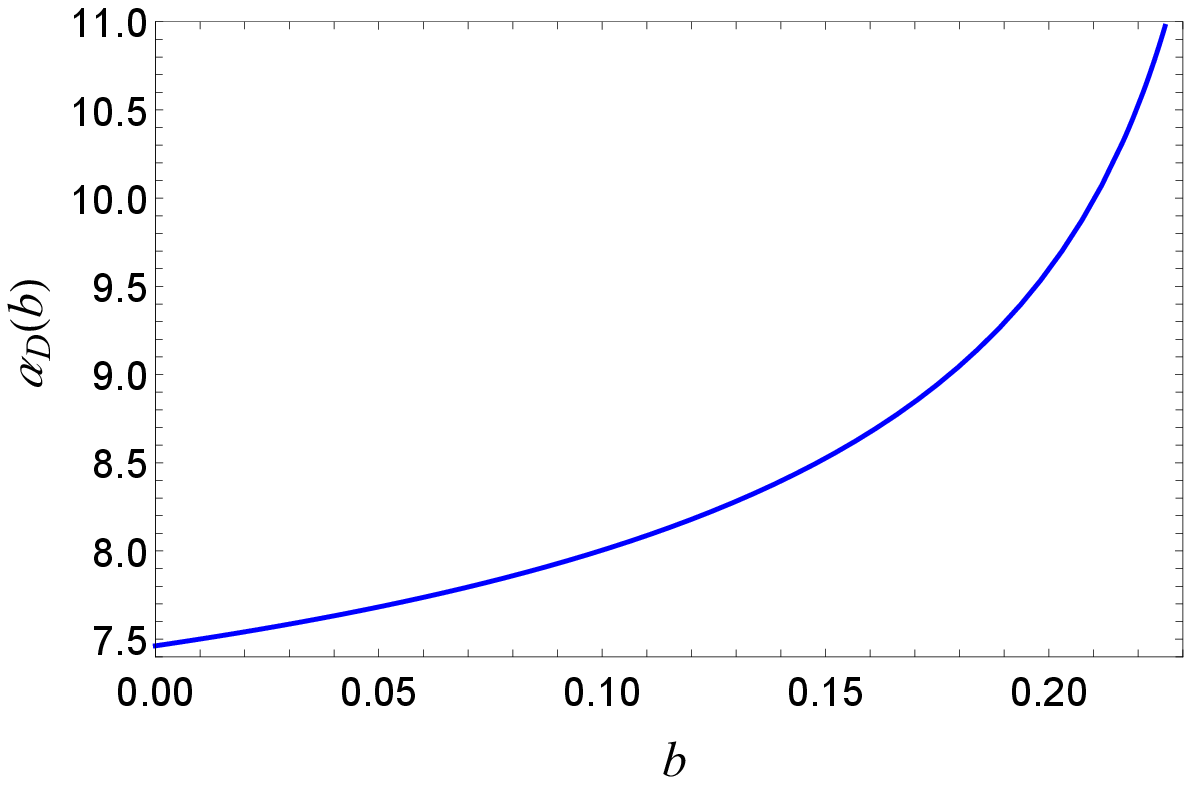}
			\end{tabular}
	\end{centering}
	\caption{(a) (left) Variation of deflection angle for REC black hole spacetime as a function of the impact parameter $u$ for different values of parameter $b$. Dots on the horizontal axis represent the values of the impact parameter $u=u_{\text{ps}}$ at which the deflection angle diverges. (b) (right) Deflection angles for REC black hole spacetime evaluated at $u=u_{\text{ps}}+0.001$ as a function of parameter $b$.} \label{fig6}		
\end{figure*}

The deflection angle, as a function of the closest approach distance $x_0$, for the spacetime Equation (\ref{metric}) reads as \citep{Virbhadra:2002ju}
\begin{align}\label{alpha}
\alpha_D(x_0)=I(x_0)-\pi= 2\int^{\infty}_{x_0}\frac{\sqrt{B(x)}dx}{\sqrt{C(x)}
\sqrt{\frac{C(x)A(x_0)}{C(x_0)A(x)}-1}}-\pi.
\end{align}
Next, let us define a new variable $z=1-x_0/x$ \citep{Bozza:2002zj,Chen:2009eu,Kumar:2021cyl} and using the relation between the impact parameter $u$ and closest approach distance $x_0$, the deflection angle as a function of the impact parameter $u$ in strong field limit yields
\begin{equation}\label{def}
\alpha_D(u)=-\bar{a}\log{\bigg(\frac{u}{u_{\text{ps}}}-1\bigg)}+\bar{b} + \mathcal{O}(u-u_{\text{ps}}) \log(u-u_{\text{ps}}),
\end{equation}
where $u \approx \theta D_{\text{OL}}$. $\bar{a}$ and $\bar{b}$  are the lensing coefficients as depicted in Figure \ref{fig5}, the calculations for which are given in \citep{Bozza:2002zj}. The deflection angle $\alpha_D(u)$ as a function of $u$ for different $b$ is depicted in Figure \ref{fig6}.

The deflection angle for REC black holes (Equation (\ref{metric})) (see Figure \ref{fig6}) is monotonically decreasing with impact parameter $u$ and deflection angle $\alpha_D(u) \to \infty$ as $u \to u_{\text{ps}}$. Interestingly, the deflection angle for REC black holes decreases with the increasing value of the parameter $b$ whereas for an impact parameter close to the critical impact parameter (see Figure \ref{fig6}), the deflection angle $\alpha_D(b) $ increases with $b$. The two infinite sets of relativistic images correspond, respectively, to clockwise and counterclockwise winding around the black hole. When $x_0 \approx x_{\text{ps}}$, the leading term of the divergence  is $z^{-1}$~\citep{Bozza:2002zj}; thus, the integral  diverges logarithmically. The coefficient $\bar{a}$ increases  while $\bar{b}$ decreases with the increasing value of $b$ (see Figure \ref{fig5}) and $\bar{a}=1$ and $\bar{b}=-0.4002$ \citep{Bozza:2002zj,Islam:2020xmy} correspond to the case of a Schwarzschild black hole (see Table~\ref{table1}).

\begin{table}
\begin{tabular}{c c c c }  
\hline\hline
\multicolumn{1}{c}{}&
\multicolumn{2}{c}{Lensing Coefficients}&
\multicolumn{1}{c}{}\\
{$b$} & {$\bar{a}$}&{$\bar{b}$} & {$u_{\text{ps}}/R_s$}\\ \hline
\hline                 
                    0  & 1.0000  & -0.40023  & 2.59808\\
\hline                  
                  0.05 & 1.03352 & -0.39609 & 2.48181\\           
\hline                 
                  0.1 & 1.08253 & -0.39889 & 2.35072\\
\hline                  
                  0.15 & 1.16297 & -0.423537 & 2.19784\\
\hline                  
                  0.2 & 1.33445 & -0.551714 & 2.00694\\ 
\hline
                  0.22 & 1.49388 & -0.758393 & 1.90995 \\
\hline\hline
	\end{tabular}
	
\caption{Estimates for the strong Lensing Coefficients $\bar{a}$, $\bar{b}$ and the critical impact parameter $u_{\text{ps}}/R_{s}$ for REC Black Hole Spacetime. The values for $b=0$ correspond to a Schwarzschild black hole.  
\label{table1}}
\end{table}  

We assume that the source and observer are far from the black hole (lens) and they are perfectly aligned. The asymptotically flat lens equation reads as \citep{Bozza:2001xd}
\begin{equation}\label{lenseq}
\beta=\theta-\frac{D_{LS}}{D_{OS}}\Delta\alpha_{n},
\end{equation}
where  $\Delta\alpha_{n}=\alpha-2n\pi$ is the offset of deflection angle looping over $2n\pi$ and $n$ is an integer. Here, $\beta$ is the angular position of the source  while $\theta$ is the angular position of the image from the optic axis. The distance between the observer and the lens and  between the observer and the source are $D_{\text{OL}}$ and $D_{OS}$, respectively. Using Equations (\ref{def}) and (\ref{lenseq}), the position of the $n$th relativistic image can be approximated as \citep{Bozza:2002zj} 
\begin{equation}\label{LensEq2}
\theta_n=\theta^0_n+\frac{u_{\text{ps}}e_n(\beta-\theta^0_n)D_{OS}}{\bar{a}D_{LS}D_{\text{OL}}},
\end{equation}
where
\begin{equation}
e_n=\text{exp}\left(\frac{\bar{b}}{\bar{a}}-\frac{2n\pi}{\bar{a}}\right),
\end{equation}
$\theta^0_n$ are the image positions corresponding to
$\alpha=2n\pi$.

As gravitational lensing conserves surface brightness, the magnification is the quotient of the solid angles subtended by the $n$th image, and the source \citep{Virbhadra:1999nm,Bozza:2002zj,Virbhadra:2008ws}. The magnification of the $n$th relativistic image reads as \citep{Bozza:2002zj}
\begin{equation}\label{mag}
\mu_n=\left(\frac{\beta}{\theta} \;  \;\frac{d\beta}{d\theta} \right)^{-1}\Bigg|_{\theta_n ^0} = \frac{u^2_{\text{ps}}e_n(1+e_n)D_{OS}}{\bar{a}\beta D_{LS}D^2_{\text{OL}}}.
\end{equation}
The first relativistic image is the brightest one, and the magnifications decrease exponentially with $n$. The magnifications are proportional to 1/$D_{\text{OL}}^2$, which is a very small factor and thus the relativistic images are very faint, unless $\beta$ has values close to zero, i.e. nearly perfect alignment.

\begin{figure*}[t] 
	\begin{centering}
		\begin{tabular}{cc}
		    \includegraphics[scale=0.75]{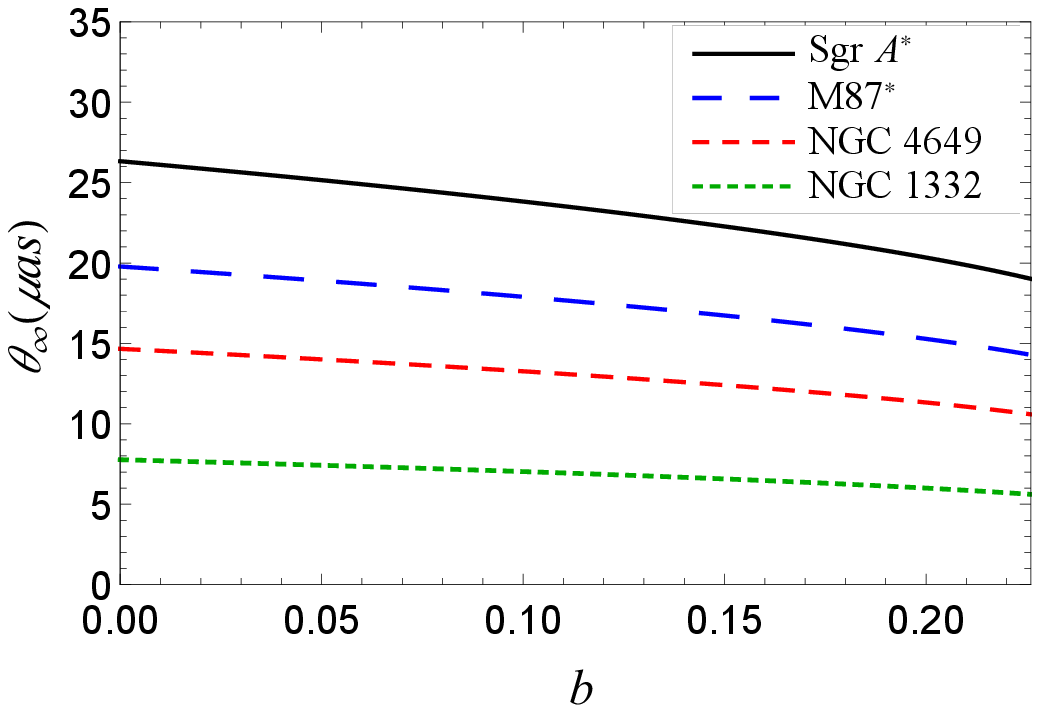}\hspace{0.5cm}
		    \includegraphics[scale=0.75]{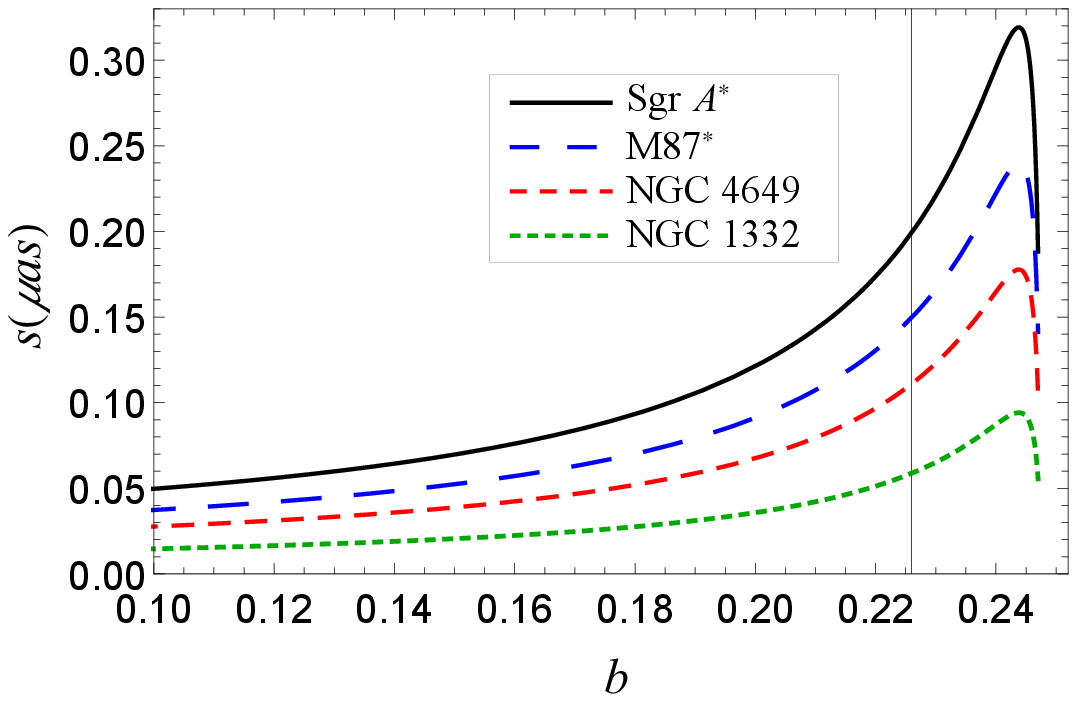}\\
			\end{tabular}
	\end{centering}
	\caption{ Behavior of lensing observables $\theta_{\infty}$ (left), $s$ (right) as a function of the parameter $b$ in strong field limit by considering that the spacetime around the compact objects at the centers of nearby galaxies is REC spacetime. The vertical black line corresponds to the extremal black hole for $b =b_{E}\approx 0.226$. For no-horizon spacetime ($b>b_{E}$), separation $s$ increases with $b$ to reach a maximum and then decreases. On the other hand, for REC black holes ($0<b<b_E$), it has an increasing behavior.}\label{fig7}
\end{figure*}

\begin{table*}
	\caption{Estimates for the Lensing Observables of Primary Images for REC Black Hole Spacetime as Compared with Schwarzschild Spacetime $(b=0)$ Considering the Supermassive Black Holes Sgr A*, M87*, NGC 4649, and NGC 1332 as a Lens. The observable $r_{\text{mag}}$ does not depend upon the mass or distance of the black hole from the observer.} \label{table2}	
\begin{ruledtabular}
\begin{tabular}{c c c c c c c c c c} 
\multicolumn{1}{c}{}&
\multicolumn{2}{c}{Sgr A*}&
\multicolumn{2}{c}{M87*}& 
\multicolumn{2}{c}{NGC 4649}&
\multicolumn{2}{c}{NGC 1332}\\
{$b$ } & {$\theta_\infty$($\mu$as)} & {$s$($\mu$as)} & {$\theta_\infty $($\mu$as)}  & {$s$($\mu$as)} & {$\theta_\infty $($\mu$as)}  & {$s$($\mu$as)} & {$\theta_\infty $($\mu$as)}  & {$s$($\mu$as)} & $r_{mag}$ \\ \hline\hline

0.0 &  26.3299  & 0.0329517 & 19.782  & 0.0247571 & 14.6615 & 0.0183488 & 7.76719 & 0.00972061 & 6.82188 \\
\hline
0.05 &  25.1516  & 0.039254 & 18.8968  & 0.0294921 & 14.0054 & 0.0218582 & 7.41962 & 0.0115797 & 6.60061 \\
\hline
0.1 &  23.8231  & 0.0496879 & 17.8987  & 0.0373313 & 13.2657 & 0.0276682 & 7.02771 & 0.0146577 & 6.3018 \\
\hline
0.15 &  22.2737  & 0.0697065 & 16.7346  & 0.0523715 & 12.4029 & 0.0388154 & 6.57066 & 0.0205631 & 5.86589 \\
\hline 
0.2 & 20.339  & 0.121317 &  15.281  & 0.0911474 & 11.3256 & 0.0675542 & 5.99993 & 0.035788 & 5.11214  \\
\hline
0.22 & 19.3561  & 0.173666  & 14.5425 & 0.130478 & 10.7783 & 0.0967039 & 5.70997 & 0.0512306 & 4.56657 \\
	\end{tabular}
	\end{ruledtabular}
\end{table*}
Having obtained the deflection angle (Equation (\ref{def})) and lens equation (Equation (\ref{lenseq})), we calculate three observables of relativistic images (see Table~\ref{table2}, Figures~\ref{fig7} and \ref{fig8}), the angular position of the asymptotic relativistic images ($\theta_{\infty}$), the angular separation between the outermost and asymptotic relativistic images ($s$) (see Figure~\ref{fig7}) and relative magnification of the outermost relativistic image with other relativistic images ($r_{\text{mag}}$) \citep{Bozza:2002zj,Islam:2021ful}
\begin{align}
\theta_{\infty}& = \frac{u_{\text{ps}}} {D_{\text{OL}}},\\
s& = \theta_1-\theta_{\infty}=\theta_\infty ~\text{exp}\left({\frac{\bar{b}}{\bar{a}}-\frac{2\pi}{\bar{a}}}\right),\\
r_{\text{mag}}& = \frac{5\pi}{\bar{a}~\text{log}(10)}\label{mag1}.
\end{align}
The strong deflection limit coefficients $\bar{a}$, $\bar{b}$ and the critical impact parameter $u_{\text{ps}}$ can be obtained after measuring $s$, $r_{\text{mag}}$ and $\theta_{\infty}$. If $\theta_{\infty}$ represents the asymptotic position of a set of images in the limit $n\rightarrow \infty$, we consider that only  the outermost image $\theta_1$ is resolved as a single image and all the remaining ones are packed together at $\theta_{\infty}$. The values obtained from measurements can be compared with those predicted by the theoretical models to check the nature of the black hole.

\begin{figure} 
		      \includegraphics[scale=0.85]{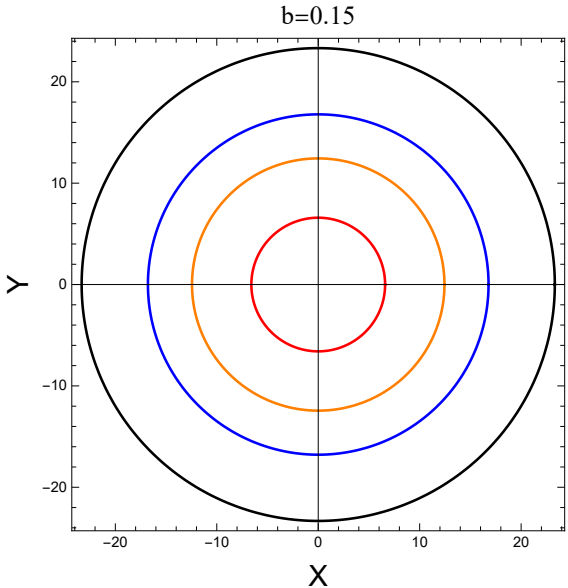}
	\caption{Einstein rings for different compact objects at the centers of nearby galaxies by considering them to be REC black holes. Black, blue, orange and red rings correspond to Sgr A*, M87*, NGC 4649 and NGC 1332. }\label{figEring}
\end{figure}

\begin{figure} 
		      \includegraphics[scale=0.85]{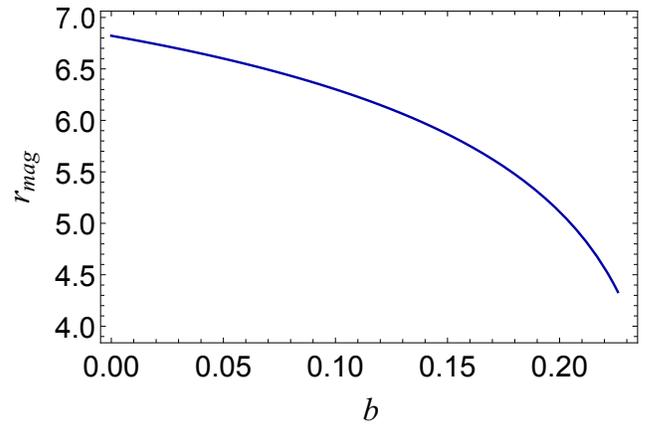}
	\caption{Behavior of strong lensing observable  $r_{\text{mag}}$, for REC black hole spacetime as a function of the parameter $b$. It is independent of the black hole's mass or distance from the observer.}\label{fig8}
\end{figure}

\begin{figure} 
		   \includegraphics[scale=0.8]{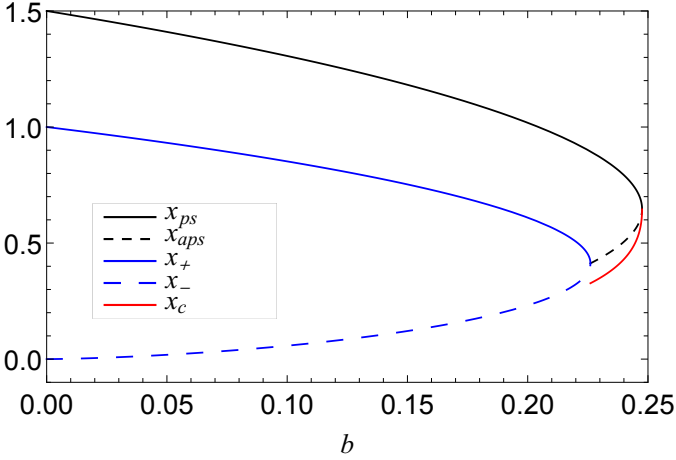}
	\caption{Locations of photon sphere $x_{\text{ps}}$ (black solid line), anti-photon sphere $x_{\text{aps}}$ (black dashed line), event horizon $x_{+}$ (blue solid line), Cauchy horizon $x_{-}$ (blue dashed line) and smaller positive root of $V_{\text{eff}}$ for $u=u_{\text{ps}}$ (red solid curve) for  REC no-horizon spacetime.}\label{fig9}
\end{figure}

\begin{figure*} 
	\begin{centering}
		\begin{tabular}{c c}
     \includegraphics[scale=0.8]{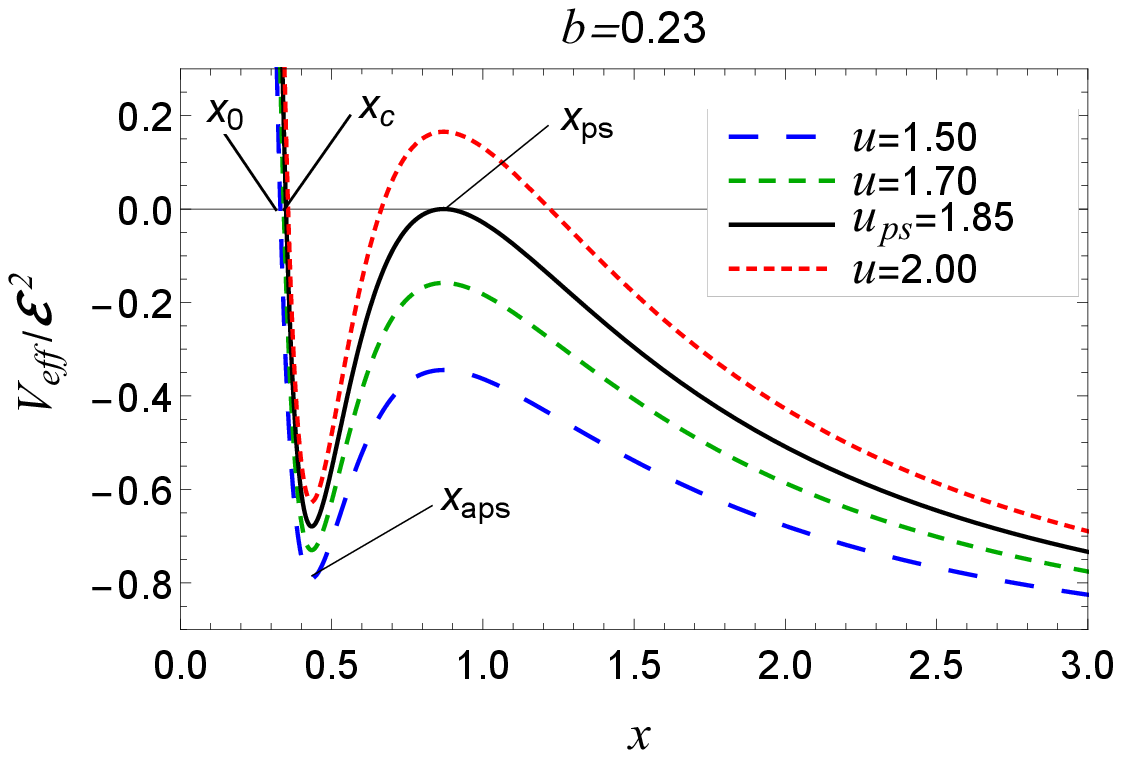}\hspace{0.5cm}
	\includegraphics[scale=0.7]{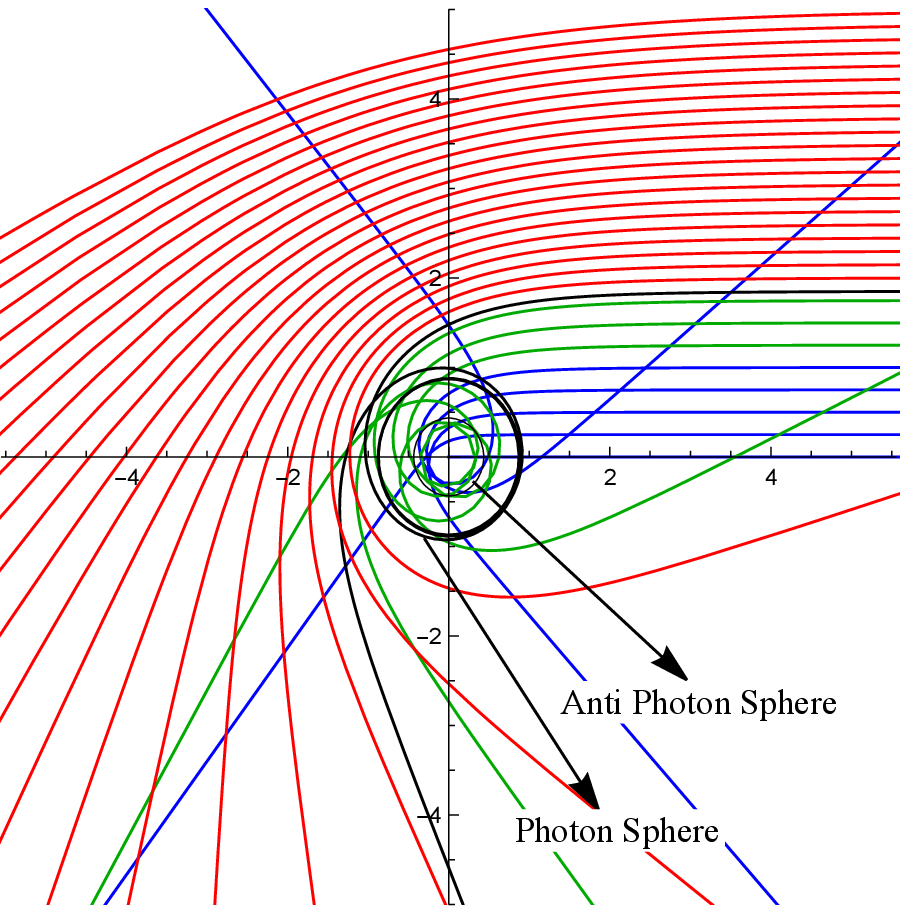}
			\end{tabular}
	\end{centering}
\caption{(a) (left) $V_{\text{eff}}$ for REC no-horizon spacetime ($b=0.23$). $x_{\text{ps}}$ and $x_{\text{aps}}$ are the locations of the photon sphere and anti-photon sphere respectively. For a photon, that has the impact parameter $u<u_{\text{ps}}$, the closest approach distance $x_0$ is close to $x_c$, where $x_c$ is the smaller positive root of $V_{\text{eff}}$ for $u=u_{\text{ps}}$. (b) (right) Trajectories of photons for REC no-horizon spacetime with different impact parameters $u$. Photons with impact parameter $u>u_{\text{ps}}$ (red curves) are scattered to infinity after reaching the closest approach distance $x_0$, which is greater than $x_{\text{ps}}$. At the same time, the photons having impact parameter $u<u_{\text{ps}}$ (green and blue curves) enter inside the photon sphere and are scattered to infinity after reaching the closest approach distance $x_0$, which is close to $x_{c}$. A photon with an impact parameter almost close to the critical impact parameter $u \approx u_{\text{ps}}$ (black curve) makes several loops around the centre before being scattered to infinity. The inner black circular curve shows the location of the anti-photon sphere.}\label{veffN}
\end{figure*}

\begin{figure} 
		   \includegraphics[scale=0.8]{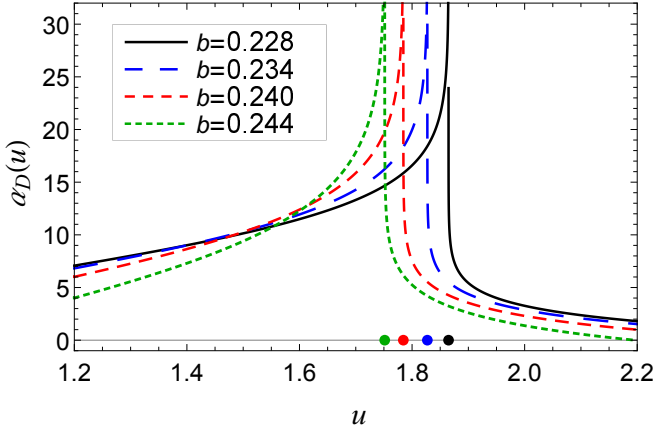}
\caption{Deflection angle for REC no-horizon spacetime as a function of the impact parameter $u$.
Dots on the horizontal axis represent the values of the critical impact parameter $u_{\text{ps}}$ at which the deflection angle diverges. The curves on the left of the dots represent strong lensing from the inside of the photon sphere ($u<u_{\text{ps}}$), while the curves on the right represent strong lensing from the outside of the photon sphere ($u>u_{\text{ps}}$).}\label{fig12}
\end{figure}

\begin{figure} 
		   \includegraphics[scale=0.8]{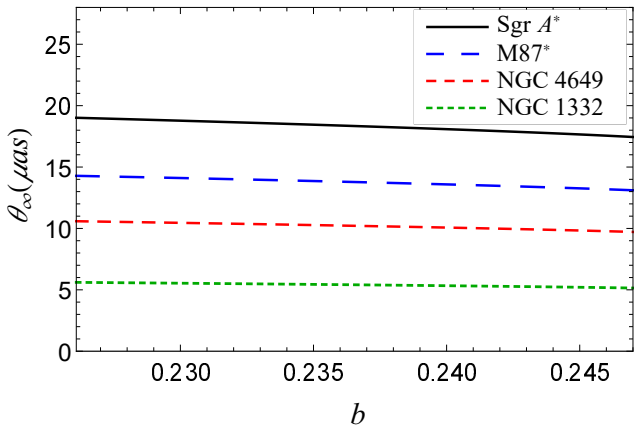}
	\caption{Behavior of lensing observable $\theta_{\infty}$ for REC no-horizon spacetime as a function of parameter $b$}\label{fig13}
\end{figure}

\begin{figure*}[t] 
	\begin{centering}
		\begin{tabular}{cc}
		 \includegraphics[scale=0.75]{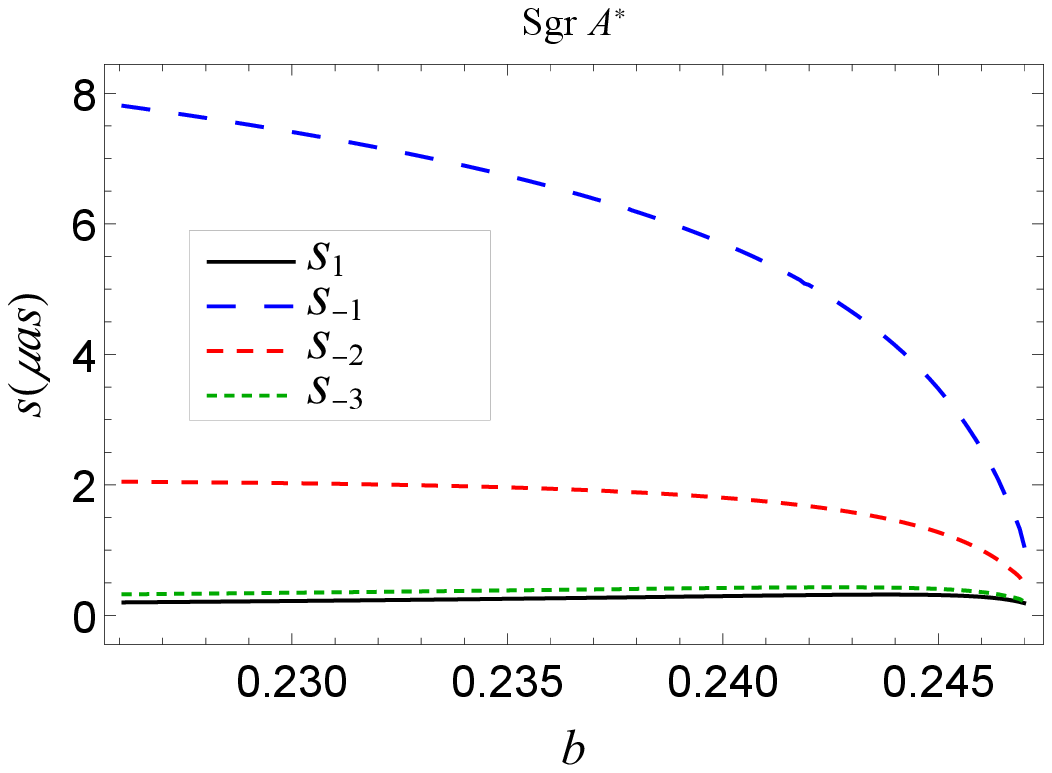}\hspace{0.5cm}
		 \includegraphics[scale=0.75]{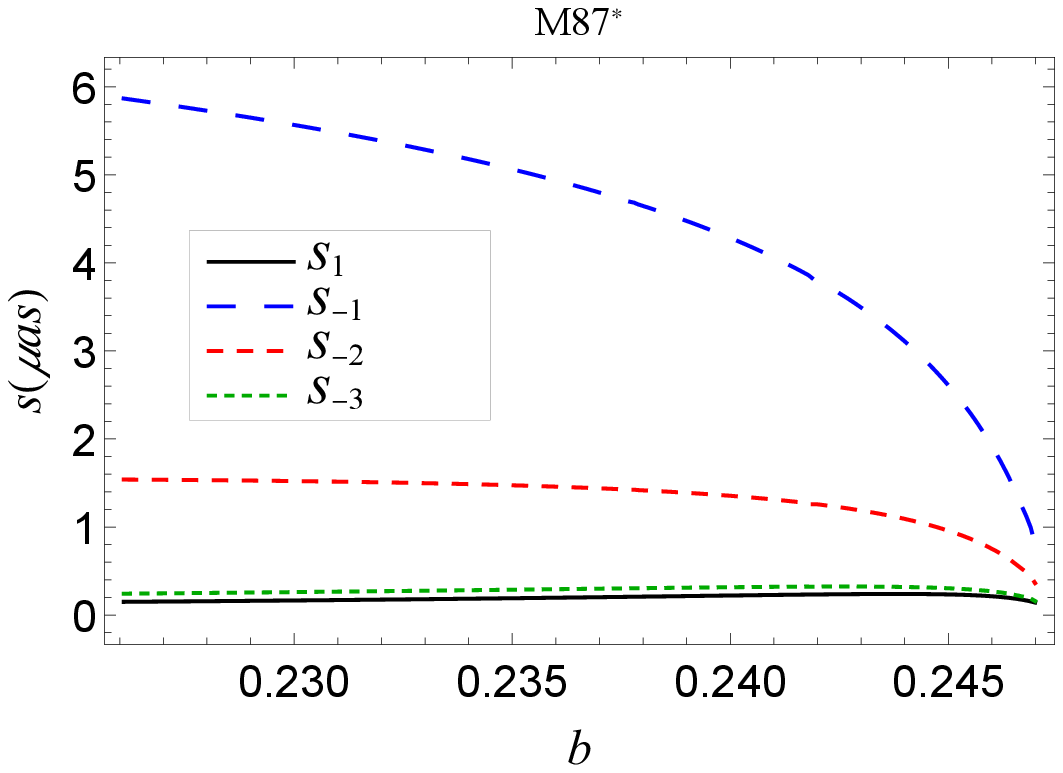}\\
		 \includegraphics[scale=0.75]{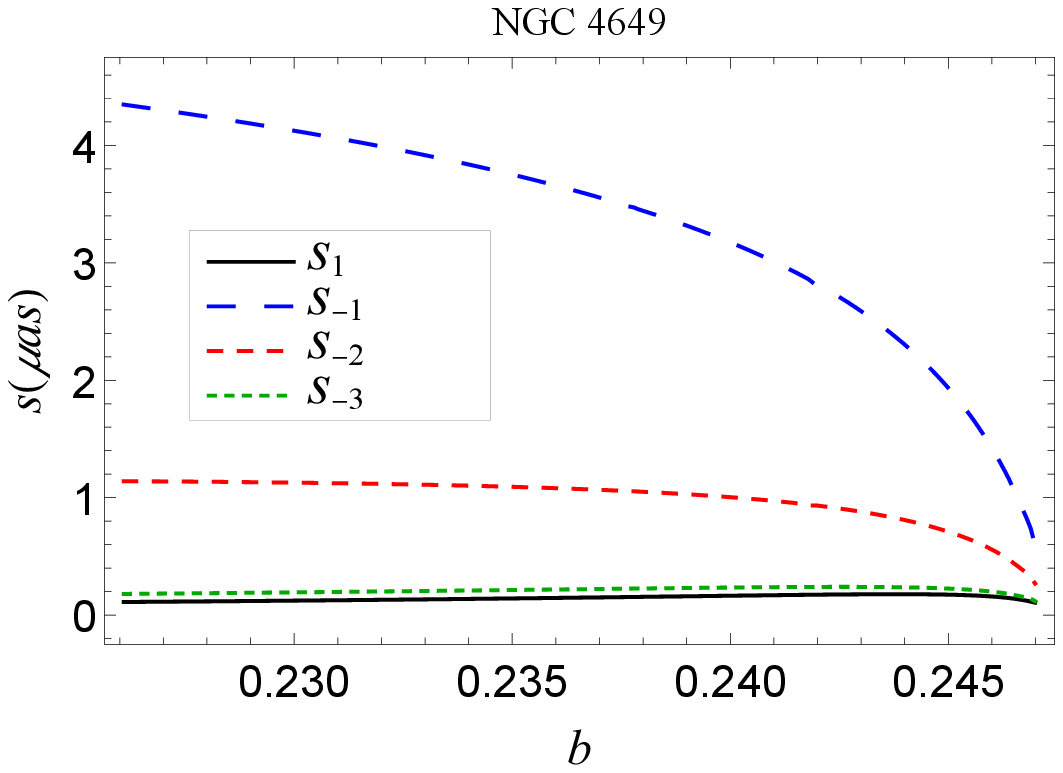}\hspace{0.5cm}
		 \includegraphics[scale=0.75]{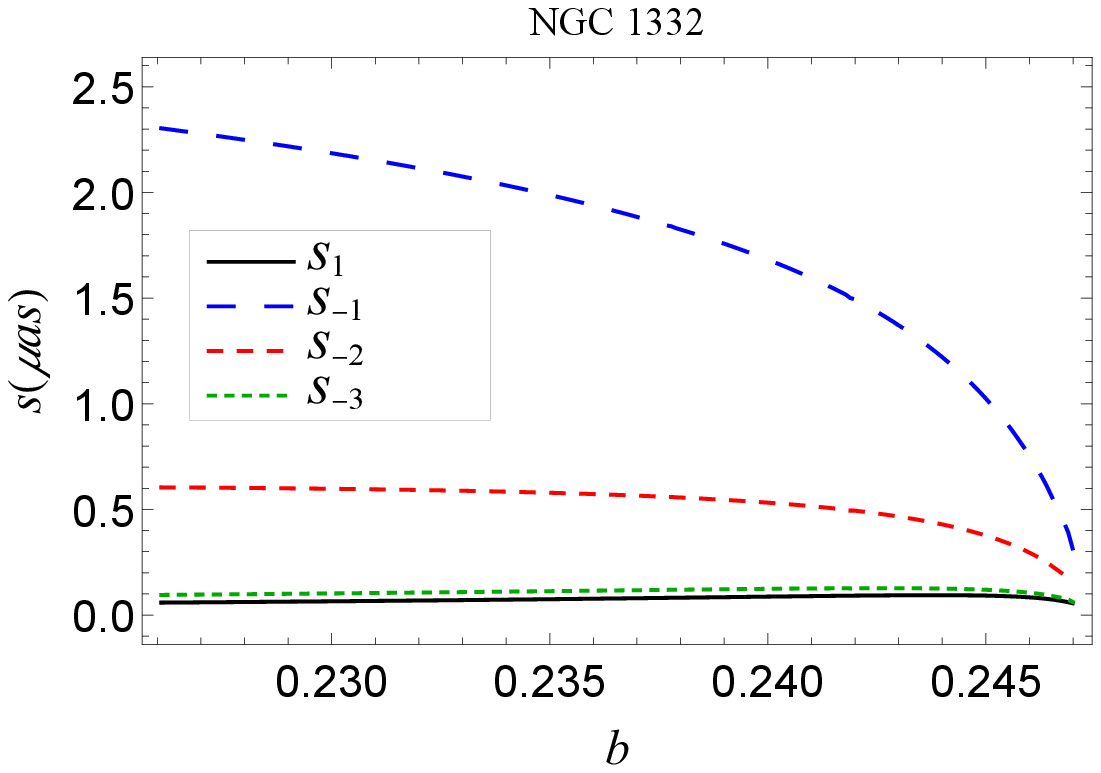}
			\end{tabular}
	\end{centering}
	\caption{Behavior of lensing observables $S_1$, $S_{-1}$, $S_{-2}$ and $S_{-3}$ for REC no-horizon spacetime as a function of parameter $b$.}\label{fig14}
\end{figure*}

\begin{figure} 
		   \includegraphics[scale=0.8]{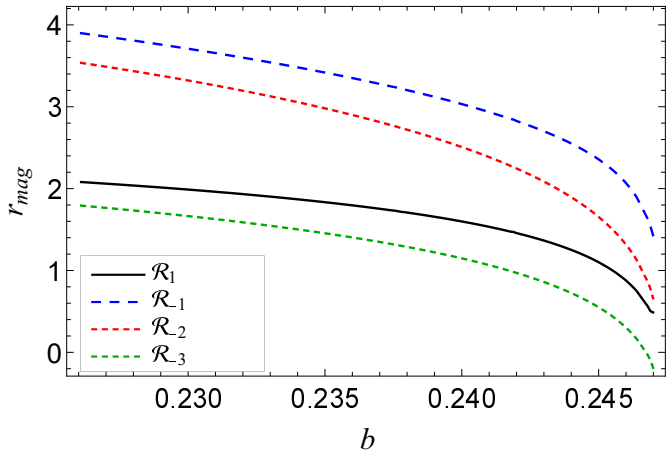}
	\caption{Behavior of strong lensing observables $\mathcal{R}_1$, $\mathcal{R}_{-1}$, $\mathcal{R}_{-2}$ and $\mathcal{R}_{-3}$ for REC no-horizon spacetime as a function of parameter $b$.}\label{fig15}
\end{figure}

\begingroup
\begin{table*}[tbh!]
	\caption{ Estimation of time delay for supermassive black holes at the center of nearby galaxies considering them as representing Schwarzschild and REC black hole spacetime $(b=0.2)$. Mass ($M$) and distance ($D_{\text{OL}}$) are given in units of solar mass and megaparsec, respectively. Time Delays are expressed in minutes.
	}\label{table3}
\begin{ruledtabular}
\begin{tabular}{c c c c c c}  
Galaxy   &           $M( M_{\odot})$      &          $D_{\text{OL}}$ (Mpc)   &     $M/D_{\text{OL}}$ & $\Delta T^s_{2,1}(\text{Schw.})$&$\Delta T^s_{2,1}(\text{REC})$           \\
\hline
Milky Way& $  4.3\times 10^6	 $ & $0.0083 $ &       $2.471\times 10^{-11}$ & $11.4968 $ &  $8.88091 $     \\
	
M87&$ 6.5\times 10^{9} $&$ 16.68 $
&$1.758\times 10^{-11}$& $17378.8 $ &  $12701.8$\\			
		
NGC 4472 &$ 2.54\times 10^{9} $&$ 16.72 $
&$7.246\times 10^{-12}$& $6791.11$ &  $5245.94$\\
			
NGC 1332 &$ 1.47\times 10^{9} $&$22.66  $
&$3.094\times 10^{-12}$& $3930.29$ &  $3036.03$\\
		
NGC 4374 &$ 9.25\times 10^{8} $&$ 18.51 $
&$2.383\times 10^{-12}$& $2473.14$ &  $1910.43$\\
			
NGC 1399&$ 8.81\times 10^{8} $&$ 20.85 $
&$2.015\times 10^{-12}$& $2355.5$ &  $1819.55$\\
			 
NGC 3379 &$ 4.16\times 10^{8} $&$10.70$
&$1.854\times 10^{-12}$& $1112.25$ &  $859.177$\\
			
NGC 4486B &$ 6\times 10^{8} $&$ 16.26 $
&$1.760\times 10^{-12}$ & $ 1604.2$ &  $ 1239.2 $\\
		
NGC 1374 &$ 5.90\times 10^{8} $&$ 19.57 $ &$1.438\times 10^{-12}$& $1577.46$ &  $1218.54$\\
			    
NGC 4649&$ 4.72\times 10^{9} $&$ 16.46 $
&$1.367\times 10^{-12}$& $ 12619.7$ &  $ 9748.35 $\\
		
NGC 3608 &$  4.65\times 10^{8}  $&$ 22.75  $ &$9.750\times 10^{-13}$& $1243.26$ &  $960.378$\\
		
NGC 3377 &$ 1.78\times 10^{8} $&$ 10.99$
&$7.726\times 10^{-13}$ & $475.913$ &  $367.629$\\
		
NGC 4697 &$  2.02\times 10^{8}  $&$ 12.54  $ &$7.684\times 10^{-13}$& $540.081$ &  $417.196$\\
			 
NGC 5128 &$  5.69\times 10^{7}  $& $3.62   $ &$7.498\times 10^{-13}$& $152.132$ &  $117.517$\\
			
NGC 1316&$  1.69\times 10^{8}  $&$20.95   $ &$3.848\times 10^{-13}$& $451.85 $ &  $349.041$\\
			
NGC 3607 &$ 1.37\times 10^{8} $&$ 22.65  $ &$2.885\times 10^{-13}$& $366.292 $ &  $282.95$\\
			
NGC 4473 &$  0.90\times 10^{8}  $&$ 15.25  $ &$2.815\times 10^{-13}$& $240.63$ &  $185.88$\\
			
NGC 4459 &$ 6.96\times 10^{7} $&$ 16.01  $ &$2.073\times 10^{-13}$ & $186.087 $ &  $143.747$\\
		
M32 &$ 2.45\times 10^6$ &$ 0.8057 $
&$1.450\times 10^{-13}$ & $6.55048 $ &  $5.06006 $    \\
			
NGC 4486A &$ 1.44\times 10^{7} $&$ 18.36  $ &$3.741\times 10^{-14}$ & $38.5008$ &  $29.7407$\\
			 
NGC 4382 &$  1.30\times 10^{7}  $&$ 17.88 $  &$3.468\times 10^{-14}$& $34.7577 $ &  $26.8493$\\
		
CYGNUS A &$  2.66\times 10^{9}  $&$ 242.7 $  &$1.4174\times 10^{-15}$& $7111.95 $ &  $5493.78$\\
		\end{tabular}
	\end{ruledtabular}
\end{table*}
\endgroup

\subsection{Einstein ring}
When a source in front of the lens can generate relativistic images and Einstein rings \citep{Luminet:1979nyg}, and the gravitational field leads to an Einstein ring if the source, lens, and observer are flawlessly aligned. However, it is adequate that just one source point is flawlessly aligned to build a complete relativistic Einstein ring \citep{Bozza:2004kq}. Thus, if the source, lens, and observer are aligned (such that $\beta=0$), then the  Equation (\ref{LensEq2}) accepts the form 
\begin{eqnarray}\label{Ering}
\theta_n^{E} &=& \left(1-\frac{D_{OS}}{D_{LS}}\frac{u_{\text{ps}} e_n}{D_{\text{OL}}\bar{a}} \right) \theta_n^{0},
\end{eqnarray} 
 which solves to give the radii of the Einstein rings. Equation (\ref{Ering}) for a special case when  the lens is equidistant between the source and observer, yields
\begin{eqnarray}\label{Ering2}
\theta_n^{E} &=& \left(1-\frac{2 u_{\text{ps}} e_n}{D_{\text{OL}}\bar{a}} \right) \left(\frac{u_{\text{ps}}}{D_{\text{OL}}}(1+e_n)\right).
\end{eqnarray} 
Since $D_{\text{OL}} \gg u_m$, the Eq.~(\ref{Ering2}) yields 

\begin{eqnarray}\label{Ering3}
\theta_n^{E} &=& \frac{u_{\text{ps}}}{D_{\text{OL}}}\left(1+e_n \right),
\end{eqnarray} 
which gives the radius of the $n$th relativistic Einstein ring. Note that $n=1$ portrays the outermost ring, and as $n$ increases, the radius of the ring decreases. Also, it can be conveniently specified from Eq.~(\ref{Ering3}) that the Einstein ring's radius increases with the black hole's mass and declines as the distance between the observer and lens increases. In Figure~\ref{figEring},  we plot the outermost relativistic Einstein rings of Sgr A*, M87* NGC 4649, and NGC 1332.

\section{Time Delay in Strong field limit}\label{sec4}
The time difference is caused by the photon taking different paths while winding the black hole, so there is a time delay between different images. If we can distinguish the time signals of the first image and other packed images, we can calculate their time delay \citep{Bozza:2003cp}. The time spent by the photon to wind around the black hole is  \citep{Bozza:2003cp}
\begin{equation}\label{TD1}
\tilde{T}(u) = \tilde{a} \log\left(\frac{u}{u_{\text{ps}}} -1\right) + \tilde{b} +\mathcal{O}(u-u_{\text{ps}}).  
\end{equation}

The images are highly demagnified, and the separation between the images is of the order of microarcseconds, so we must at least distinguish the outermost relativistic image from the rest, and we assume the source to be variable, which generally are abundant in all galaxies, otherwise, there is no time delay to measure.  For spherically symmetric black holes, the time delay between the first and second relativistic image reads as
\begin{equation}\label{deltaT}
\Delta T^s_{2,1} = 2\pi u_{\text{ps}} = 2\pi D_{\text{OL}} \theta_{\infty}.
\end{equation}  
Using Eq.~(\ref{deltaT}), if we can measure the time delay with an accuracy of $ 5\% $ and the critical impact parameter with negligible error, we can obtain the distance of the black hole with an accuracy of $5\%$. In Table~\ref{table3}, we compare $\Delta T^s_{2,1}$ for the supermassive black holes at the center of several galaxies representing Schwarzschild and REC black hole spacetime.

\begin{table*}[t]
\centering
	\caption{
		{\bf Magnifications  of first-order and second-order relativistic images due to lensing by Sgr A* with $d=D_{\text{LS}}/D_{\text{OS}}=0.5$}: Schwarzschild and REC black hole spacetime $(b=0.2)$ predictions for magnifications $\mu_n$  are given for different values of angular source position $\beta$. {\bf (a)} $1p$ and $1s$ refer to first-order relativistic images on the same side as primary and secondary images, respectively. {\bf (b)}  We have used $M_{\text{Sgr A*}}= 4.3\times 10^6	\, {\rm m}$, $D_{\text{OL}}= 8.35 \times 10^{3} \, {\rm pc}$ {\bf (c)} Angular positions of first-order relativistic images for Schwarzschild and REC black hole spacetime are, respectively, $\theta_{1p,{\rm SCH}}\approx -\theta_{1s,{\rm SCH}}\approx 26.3299  \mu as$ and $\theta_{1p,{\rm DM}}\approx -\theta_{1s,{\rm DM}}\approx 20.339  \mu as$ and are highly insensitive to the angular source position $\beta$.
	}\label{table6}
\resizebox{\textwidth}{!}{%
\begin{tabular}{p{0.7cm} p{2.3cm} p{2.3cm} p{2.4cm} p{2.4cm} p{2.3cm} p{2.3cm} p{2.4cm} p{2.3cm}}
\hline\hline
                                                        &
			\multicolumn{4}{c}{REC black hole spacetime}&
			\multicolumn{4}{c}{Schwarzschild spacetime}\\
    $\beta$($\mu${\em as}) &  $\mu_{1p,{\rm DM}} $ & $ \mu_{2p,{\rm DM}}$&$\mu_{1s,{\rm DM}}$&$\mu_{2s,{\rm DM}}$&$\mu_{1p,{\rm SCH}}$&$\mu_{2p,{\rm SCH}}$&$\mu_{1s,{\rm SCH}}$&$\mu_{2s,{\rm SCH}}$\\
	\hline 	\hline

		
	    	$10^{0}  $&$1.80359 \times 10^{-11}$&$ 1.61705  \times	10^{-13} $&$-1.80359 \times 10^{-11} $&$ -1.61705  \times	10^{-13} $&$8.52495 \times 10^{-12}  $&$1.59  \times	10^{-14} $&$-8.52495 \times 10^{-12}  $&$-1.59  \times	10^{-14}  $\\
			
			$10^{1}  $&$1.80359 \times 10^{-12}$&$ 1.61705  \times	10^{-14} $&$-1.80359\times 10^{-12} $&$ -1.61705  \times	10^{-14} $&$8.52495 \times 10^{-13}$&$1.59  \times	10^{-15} $&$-8.52495 \times 10^{-13}  $&$- 1.59  \times	10^{-15}	$\\
			
			$10^{2}  $&$ 1.80359 \times 10^{-13} $&$ 1.61705 \times	10^{-15} $&$-1.80359 \times 10^{-13} $&$ -1.61705  \times	10^{-15} $&$8.52495 \times 10^{-14} $&$1.59  \times	10^{-16} $&$-8.52495 \times 10^{-14} $&$ -1.59  \times	10^{-16}	$\\
			
			$10^{3}  $&$1.80359 \times 10^{-14} $&$ 1.61705  \times	10^{-16} $&$-1.80359 \times 10^{-14}$&$ -1.61705  \times	10^{-16} $&$8.52495 \times 10^{-15}  $&$ 1.59  \times	10^{-17} $&$-8.52495 \times 10^{-15}  $&$ -1.59  \times	10^{-17} 		$\\
			
			$10^{4}  $&$1.80359 \times 10^{-15} $&$  1.61705  \times	10^{-17} $&$-1.80359 \times 10^{-15} $&$ -1.61705  \times	10^{-17} $&$8.52495 \times 10^{-16} $&$1.59  \times	10^{-18}$&$-8.52495 \times 10^{-16}  $&$ -1.59  \times	10^{-18} 	$\\
\hline\hline
\end{tabular}
}
\end{table*}

\begin{table*}[t]
\centering
	\caption{
		{\bf Magnifications  of first- and second-order relativistic images due to lensing by M87* with $d=D_{\text{LS}}/D_{\text{OS}}=0.5$}: Schwarzschild and REC black hole spacetime $(b=0.2)$ predictions for magnifications $\mu_n$ are given for different values of angular source position $\beta$. {\bf (a)} $1p$ and $1s$ refer to first-order relativistic images on the same side as primary and secondary images, respectively. {\bf (b)}  We have used $M_{\text{M87*}}=  6.5 \times 10^9 \, {\rm m}$, $D_{\text{OL}}=  16.8  \times 10^6 \, {\rm pc}$. {\bf (c)} Angular positions of first-order relativistic images for Schwarzschild and REC black hole spacetime are, respectively, $\theta_{1p,{\rm SCH}}\approx -\theta_{1s,{\rm SCH}}\approx 19.8068  \mu as$ and $\theta_{1p,{\rm DM}}\approx -\theta_{1s,{\rm DM}}\approx 15.3722 \mu as$ and are highly insensitive to the angular source position $\beta$.
	}\label{table7}
\resizebox{\textwidth}{!}{%
\begin{tabular}{p{0.7cm} p{2.3cm} p{2.3cm} p{2.4cm} p{2.4cm} p{2.3cm} p{2.3cm} p{2.4cm} p{2.4cm}}
\hline\hline
	                                            &
	\multicolumn{4}{c}{REC black hole spacetime}&
	\multicolumn{4}{c}{Schwarzschild spacetime}\\
\hline			
     $\beta$($\mu${\em as})&  $\mu_{1p,{\rm DM}} $ & $ \mu_{2p,{\rm DM}}$&$\mu_{1s,{\rm DM}}$&$\mu_{2s,{\rm DM}}$&$\mu_{1p,{\rm SCH}}$&$\mu_{2p,{\rm SCH}}$&$\mu_{1s,{\rm SCH}}$&$\mu_{2s,{\rm SCH}}$\\
	\hline 	\hline

		
			$10^{0}$  & $1.01808 \times 10^{-11}$ & $9.12786  \times	10^{-14}$ &$-1.01808  \times 10^{-11} $&$ -9.12786  \times	10^{-14} $&$4.75466  \times 10^{-12}  $&$8.86797 \times	10^{-15} $&$-4.75466 \times	10^{-12}  $&$-8.86797  \times	10^{-15}  $\\
			
			$10^{1}  $&$1.01808\times10^{-12}$&$ 9.12786  \times	10^{-15} $&$ -1.01808\times10^{-12}$&$ -9.12786  \times	10^{-15} $&$4.75466\times10^{-13}$&$8.86797 \times	10^{-16} $&$-4.75466\times10^{-13}  $&$ -8.86797  \times	10^{-16}	$\\
			
			$10^{2}  $&$1.01808\times10^{-13} $&$ 9.12786  \times	10^{-16} $&$ -1.01808\times10^{-13}$&$ -9.12786  \times	10^{-16} $&$ 4.75466\times10^{-14} $&$8.86797  \times	10^{-17} $&$-4.75466\times10^{-14} $&$ -8.86797  \times	10^{-17}	$\\
			
			$10^{3}  $&$1.01808\times10^{-14} $&$ 9.12786  \times	10^{-17} $&$-1.01808\times10^{-14}$&$-9.12786  \times	10^{-17} $&$ 4.75466\times10^{-15}  $&$ 8.86797 \times	10^{-18} $&$-4.75466 \times 10^{-15}  $&$ -8.86797  \times	10^{-18}		$\\
			
			$10^{4}  $&$1.01808\times10^{-15} $&$ 9.12786 \times	10^{-18} $&$-1.01808\times10^{-15} $&$ -9.12786  \times	10^{-18} $&$4.75466\times10^{-16} $&$8.86797 \times	10^{-19} $&$-4.75466\times10^{-16}  $&$ -8.86797 \times	10^{-19}  		$\\
\hline\hline			
\end{tabular}%
}
\end{table*}
\section{Strong gravitational lensing in REC no-horizon spacetimes}\label{sec5}
Next, we investigate gravitational lensing in the strong deflection limit by REC no-horizon spacetime with $b_E< b \leq b_P $. Since there is no horizon, photons with impact parameter $u<u_{\text{ps}}$ need not necessarily fall to the center and as there exists a potential barrier near $x_c$ (see Figures   \ref{fig9} and \ref{veffN}), where $x_{c}$ is the smaller positive root of $V_{\text{eff}}$ for $u=u_{\text{ps}}$, the photons will emerge after reaching the closest approach distance $x_{0}\approx x_{c}$. Thereby, the lensing will occur not only from the outside but also from the inside of the photon sphere (see Figure \ref{veffN}), resulting in two sets of infinite images, created by lensing from the outside and the inside of the photon sphere. The lensing from outside of photon sphere is exactly similar to the black hole case and the prescription given in Section \ref{sec3} remains valid.

\subsection{Lensing from inside the photon sphere}
In this case, the impact parameter $u$ of the photon is less than the critical value $u_{\text{ps}}$. Due to the presence of the potential barrier near $x_c$, the photon, after entering inside the photon sphere, gets reflected at the closest approach distance $x_0 \approx x_c$ and exits toward the $\infty$. 

The deflection angle as a function of the impact parameter $u$ is given by Shaikh et al. (\citeyear{Shaikh:2019itn}) as
\begin{equation}
\alpha_D(u)=-\bar{p}\log \left( \frac{u_{\text{ps}}^2}{u^2}-1 \right) +\bar{q} +\mathcal{O}((u_{\text{ps}}^2-u^2)\log(u_{\text{ps}}^2-u^2)),
\label{def2}
\end{equation}
where
\begin{equation}
\bar{p}=2\sqrt{\frac{2B(x_{\text{ps}}) A(x_{\text{ps}})}{C^{''}(x_{\text{ps}}) A(x_{\text{ps}})-C(x_{\text{ps}}) A^{''}(x_{\text{ps}})}},
\end{equation}
\begin{eqnarray}
\bar{q} &&= \bar{p}\log \left[2x_{\text{ps}}^2\left(\frac{C^{''}(x_{\text{ps}})}{C(x_{\text{ps}})}-\frac{A^{''}(x_{\text{ps}})}{A(x_{\text{ps}})}\right)\left(\frac{x_{\text{ps}}}{x_c}-1\right)\right] \nonumber\\&&  +I_R(x_c)-\pi.
\end{eqnarray}
Figure~\ref{fig12} shows the deflection angle for no-horizon spacetime.
Unlike in the case of a black hole, photons with the impact parameter $u<u_{\text{ps}}$ are also lensed. A photon lensed from the inside of the photon sphere is represented by the curves on the left of the dots, while photon lensing from the outside of the photon sphere is depicted by the curves on the right of the dots.  Using  Eq's.~(\ref{lenseq}) and (\ref{def2}), the angular position of the $n$th relativistic image formed from the inside of the photon sphere can be written as \citep{Shaikh:2019itn}
\begin{equation}
\theta_{-n}=\theta^0_{-n}-\frac{u_{\text{ps}}e_{-n}(\beta-\theta^0_{-n})D_{OS}}{2\bar{p}D_{LS}D_{\text{OL}}\left(1+e_{-n}\right)^\frac{3}{2}},
\end{equation}
where
\begin{equation}
\theta^0_{-n}=\frac{u_{\text{ps}}}{D_{\text{OL}}}\frac{1}{\sqrt{1+e_{-n}}},
\end{equation}
and
\begin{equation}
e_{-n}=\text{exp}\left({\frac{\bar{q}-2n\pi}{\bar{p}}}\right).
\end{equation}
The correction to $\theta^0_{-n}$ is negligible compared to $\theta^0_{-n}$. Therefore, the angular positions of the images are approximated by $\theta^0_{-n}$ in order to calculate the observables such as separation and magnifications of the images. 

We define separation $s_{-n}$ as
\begin{equation}
s_{-n}=\theta_{\infty}-\theta_{-n},   
\end{equation}
which is the angular separation between the $n$th image formed by lensing from the inside of the photon sphere and the image corresponding to the impact parameter $u=u_{\text{ps}}$. $s_1$ is defined as the angular separation between the first image formed by lensing from the outside of the photon sphere and the image corresponding to the impact parameter $u=u_{\text{ps}}$. We find that the first three images formed by lensing from the inside of the photon sphere are well separated in comparison to the images formed by lensing from the outside of the photon sphere. 
We define relative magnification $\mathcal{R}_{1}$ for the first image formed from the outside of the photon sphere as \citep{Shaikh:2019itn}
\begin{equation}
\mathcal{R}_{1}=2.5\log\frac{\mu_{1}}{\sum_{m=2}^\infty \mu_m+\sum_{m=4}^\infty |\mu_{-m}|} 
\end{equation}
and relative magnification $\mathcal{R}_{-n}$ for the first three images formed from the inside of the photon sphere as \citep{Shaikh:2019itn}
\begin{equation}
\mathcal{R}_{-n}=2.5\log\frac{|\mu_{-n}|}{\sum_{m=2}^\infty \mu_m+\sum_{m=4}^\infty |\mu_{-m}|} \quad (n=1,2,3)
\end{equation}
where
\begin{eqnarray}
\sum_{m=2}^\infty \mu_m  = && \frac{u^2_{\text{ps}}D_{OS}}{\bar{a}\beta D^2_{\text{OL}}D_{LS}}
\nonumber\\&&\frac{\text{exp}\left(\frac{\bar{b}}{\bar{a}}\right)\left[\text{exp}\left(\frac{4\pi}{\bar{a}}\right)+\text{exp}\left(\frac{2\pi}{\bar{a}}\right)+\text{exp}\left(\frac{\bar{b}}{\bar{a}}\right)\right]}{\text{exp}\left(\frac{4\pi}{\bar{a}}\right)\left[\text{exp}\left(\frac{4\pi}{\bar{a}}\right)-1\right]}\nonumber\\
\end{eqnarray}
\begin{eqnarray}
\sum_{m=4}^\infty |\mu_{-m}| = && \frac{u^2_{\text{ps}}D_{OS}}{2\bar{p}\beta D^2_{\text{OL}}D_{LS}}\nonumber\\&&
\frac{\text{exp}\left(\frac{\bar{q}}{\bar{p}}\right)\left[\text{exp}\left(\frac{8\pi}{\bar{p}}\right)+\text{exp}\left(\frac{6\pi}{\bar{p}}\right)-2\text{exp}\left(\frac{\bar{q}}{\bar{p}}\right)\right]}{\text{exp}\left(\frac{12\pi}{\bar{p}}\right)\left[\text{exp}\left(\frac{4\pi}{\bar{p}}\right)-1\right]}\nonumber\\
\end{eqnarray}

with $\mu_m$ and $\mu_{-m}$ being the fluxes of the $m$th image formed, respectively, from outside and inside of the photon sphere.

\section{Gravitational lensing Observables}\label{sec6}
We use REC spacetime to estimate  the observables of massive dark objects  in neighboring galaxies, particularly Sgr A*, M87*, NGC 4649, and NGC 1332, modeling them both as black holes as well as horizonless compact objects. A light ray traveling close to the photon sphere of a black hole or a horizonless compact object, depending on the impact parameter, passes around it once, twice, or many times before reaching an observer, creating a series of theoretically infinite number of relativistic images. We compute the angular positions as well as the separation and relative magnification of higher-order relativistic primary and secondary images, respectively, on the same and opposite sides of the source,  taking $d=D_{LS}/D_{OS}=0.5$.  

When the spacetime is considered a black hole, we discover that an REC black hole with the same mass and distance has a smaller position in the innermost image (apparent radius of the photon sphere: $\theta_{\infty}$) than the Schwarzschild black hole spacetime and it further decreases with the higher values of $b$ with $\theta_{\infty}\in~(19.0178,26.3299)~\mu$as for Sgr A* with a deviation of $7.3121~\mu$as from the Schwarzschild black hole and for M87* $\in~(14.2884,19.782)~\mu$as with a deviation of $5.4936~\mu$as. Contrarily, the the angular separation  between the outermost relativistic image and the others packaged at the photon sphere   $\in(0.0329517,0.199342)~\mu$as   for Sgr A* and $\in(0.0247571,0.149769)~\mu$as for M87*, with the separation $s$ being five times greater than that for Schwarzschild black hole in both cases. In addition to being considerably faint, the spacing is well beyond what is now possible with technology. Figure~\ref{fig7} depicts these typical observables, which comprises the position of the innermost image $\theta_{\infty}$ and the separation $s$, and are subsequently tabulated in Table \ref{table2}. Moreover, we calculated the relative magnification of the outermost relativistic image caused by REC black hole spacetime gravitational lensing and discovered that the magnification decreases with parameter  $b$ (cf.~Figure~\ref{fig8}) and $\in(4.56657,6.82188)$. The relative magnifications of first- and second-order images in REC black hole spacetime and Schwarzschild spacetime for different angular positions of the source using Eq.~(\ref{mag}) are shown in Tables~\ref{table6} and \ref{table7}. We noticed that higher-order images are significantly demagnified, as expected. 
\begin{figure*}[t]
	\begin{centering}
		\begin{tabular}{c c}
\includegraphics[scale=0.7]{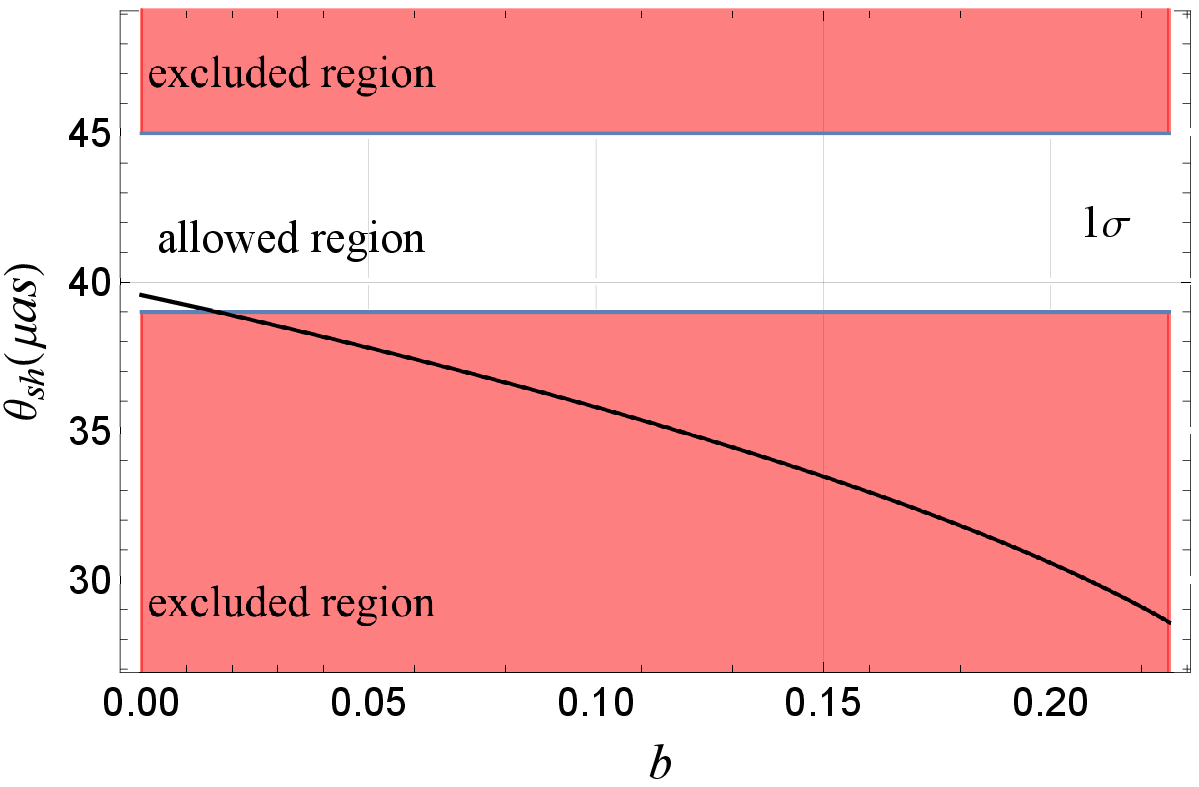}\hspace{0.5cm}
\includegraphics[scale=0.7]{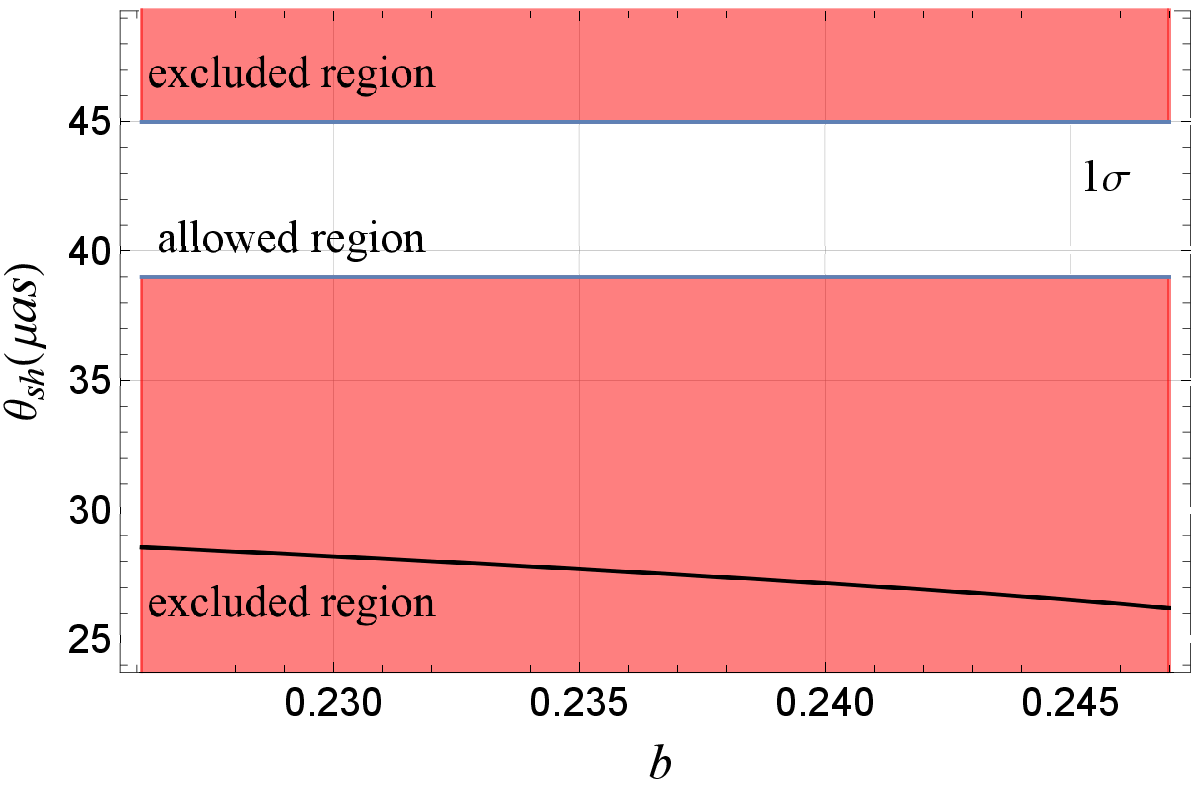}
			\end{tabular}
	\end{centering}
\caption{Shadow angular diameter $\theta_{\text{sh}}(=2\theta_{\infty})$ as a function of $b$, when M87* is modeled as REC spacetimes:(a) black hole (left) (b) no-horizon spacetime (right). The dark black line corresponds to calculated values of $2\theta_{\infty}$ as a function of $b$. The white/red  regions represent areas that are 1 $\sigma$ consistent/inconsistent with the 2017 EHT observations, indicating that the latter imposed constraints on parameter $b$.}\label{M87parameter}
\end{figure*}
Considering the no-horizon REC spacetime, we discovered a series of relativistic images outside as well as inside the photon sphere, a feature that a black hole lacks. As depicted in Figure~\ref{fig13}, $\theta_{\infty}$ decreases  with increasing value of the parameter $b$, just as in the black hole situation and  $\in~(17.4462, 19.0178)~\mu$as for Sgr A* and $\in~(13.1076,14.2884)~\mu$as for M87*. Together with the image at $\theta_1$ outside the photon sphere,  we have taken into account the three additional images at $\theta_{-1}$, $\theta_{-2}$, and $\theta_{-3}$, which are well spaced with $s_{1} \in$ (0.1882,0.319045)~$\mu$as, $s_{-1} \in$ (1.12743,7.8168)~$\mu$as,  $s_{-2} \in$ (0.512824,2.04872)~$\mu$as, and $s_{-3} \in$ (0.226226,0.322697)~$\mu$as for Sgr A* and $s_{1} \in$ (0.141397,0.239703)~$\mu$as, $s_{-1} \in$ (0.847053,5.87288)~$\mu$as,  $s_{-2} \in$ (0.385292,1.53924)~$\mu$as, and $s_{-3} \in$ (0.169967,0.242447)~$\mu$as for M87* (see Figure~\ref{fig14}). As we move from  $\theta_1$ or  $\theta_{-1}$ toward $\theta_{\infty}$, the angular separation between the consecutive images gets smaller. Further, the brightest images are those with a relative magnitude of $\mathcal{R}_{-1}$ with $\mathcal{R}_{-1} \in$ (1.42485,3.90258), $\mathcal{R}_{-2} \in$ (0.669416,3.54039), $\mathcal{R}_{-3} \in$  (-0.17369,1.79488), $\mathcal{R}_{1} \in$ (0.42613,2.0822) and  brightness decreases in a similar way as separation as we go toward $\theta_{\infty}$ from $\theta_1$ or  $\theta_{-1}$ (see Figure~\ref{fig15}). Similar to the black hole situation, the relative magnitude of the brightness of the images both inside and outside the photon sphere decreases  with  parameter $b$. 

Finally, we gather updated data from 14 supermassive black holes whose masses and distances considerably vary  in  Table~\ref{table3} in order to determine the time delays $\Delta T^s_{2,1}$ between first- and second-order relativistic primary images. The time delay  $\Delta T^s_{2,1}$, for Sgr A*, M87*, NGC 4649, and NGC 1332, respectively, can reach $\sim 8.8809$, $\sim 12701.8$, $\sim 9748.35$, and $\sim 3036.03$ minutes (see ~Table \ref{table3}), with the deviation from the Schwarzschild black hole of same mass and distance being $\sim 2.6159$, $\sim 4677$, $\sim 2871.35$, and $\sim 894.26$ minutes. As a result, the time delay in Sgr A* is much shorter for observation and much more difficult to measure. In the case of other black holes, the time delay $\Delta T^s_{2,1}$ can be in the hundreds of hours, which are adequate times for astronomical measurements, provided we have sufficient angular resolution between two relativistic images.

\section{Constraints from EHT observations of M87* and Sgr A*}\label{sec7}
In 2019, black holes became a reality  with the  EHT Collaboration  horizon-scale image of the supermassive black hole M87*  \citep{EventHorizonTelescope:2019dse,EventHorizonTelescope:2019pgp,EventHorizonTelescope:2019ggy}.  Using a distance of $d=16.8$ Mpc and their estimated mass of M87*  $M=(6.5 \pm 0.7) \times 10^9 M_\odot$, the EHT Collaboration found a compact emission region size with angular diameter $\theta_d=42\pm 3\, \mu $as  with the central flux depression with a factor of $\gtrsim 10$, which is the black hole shadow. 
Recently in 2022, the EHT observation also unveiled the image of supermassive black hole Sgr A* in our Milky Way showing a ring of diameter $\theta_d= 48.7 \pm 7\,\mu$as, inferring a black hole mass of $M = 4.0^{+1.1}_{-0.6} \times 10^6 M_\odot $ and Schwarzschild shadow deviation $\delta = -0.08^{+0.09}_{-0.09}~\text{(VLTI)},-0.04^{+0.09}_{-0.10}~\text{(Keck)}$. and provided us with yet another tool to investigate the nature of strong field gravity \citep{EventHorizonTelescope:2022xnr,EventHorizonTelescope:2022urf,EventHorizonTelescope:2022xqj}. 
The EHT result for Sgr A* also only calculated the emission ring angular diameter $\theta_d=(51.8\pm 2.3)\mu$as with the prior perceived estimates  $M$ mentioned above and $D_{LS}=8.15\pm 0.15$ kpc \citep{EventHorizonTelescope:2022xqj}. The horizon-scale images of supermassive black holes furnish a theoretically fresh avenue for testing the theories of gravity and use the EHT results of the Sgr A* and  M87* shadow size to infer constraints on the additional parameter of the underlying theory for a wide variety of black holes \citep{Kumar:2020yem,Ghosh:2020syx,Kumar:2018ple,Afrin:2021imp,EventHorizonTelescope:2021dqv,EventHorizonTelescope:2022xqj,Afrin:2021wlj,Ghosh:2022kit,Islam:2022ybr,KumarWalia:2022aop,Zakharov:2021gbg}
\begin{figure*}[t]
	\begin{centering}
		\begin{tabular}{c c}
\includegraphics[scale=0.7]{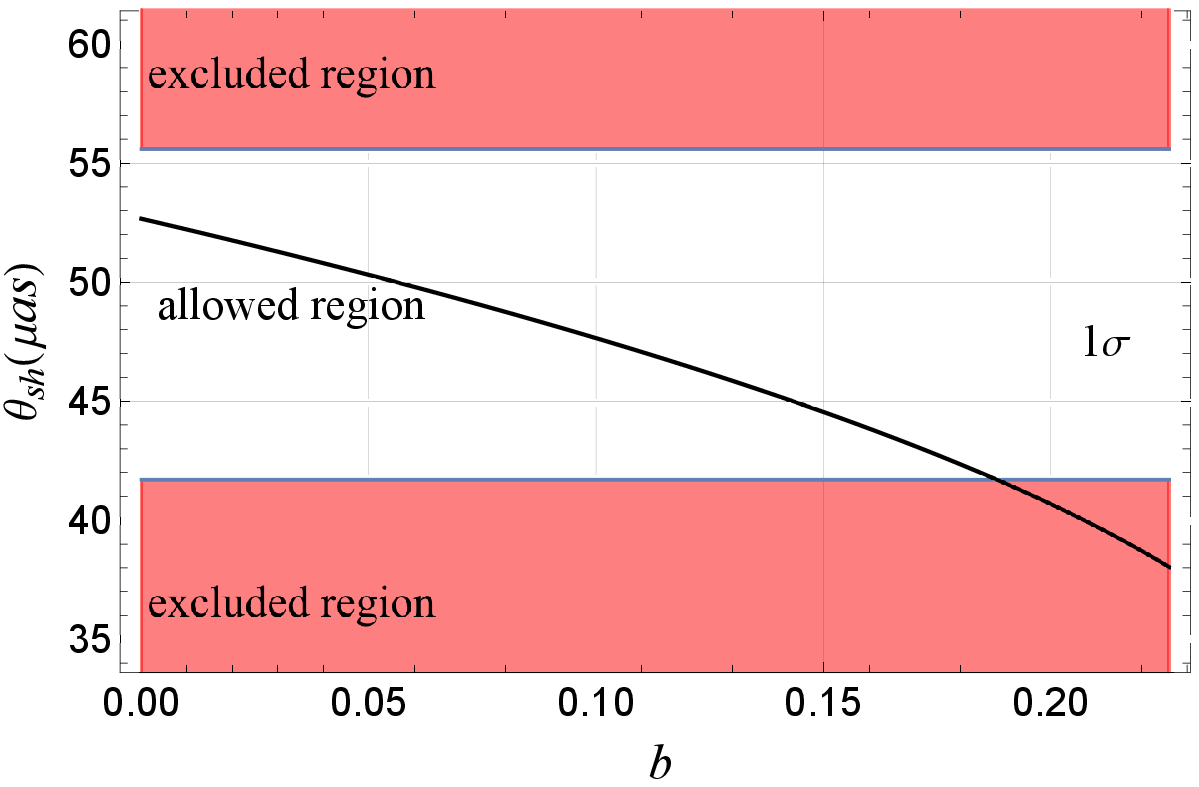}\hspace{0.5cm}
\includegraphics[scale=0.7]{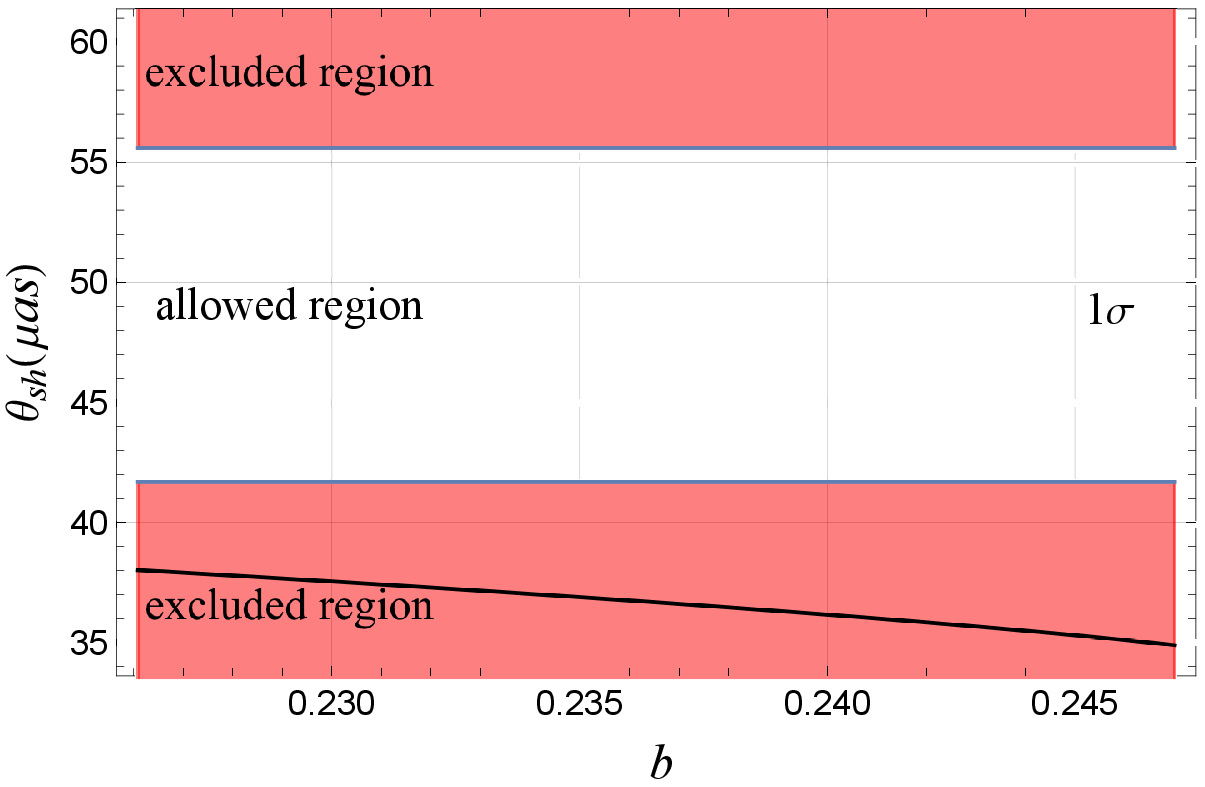}
			\end{tabular}
	\end{centering}
\caption{Shadow angular diameter $\theta_{\text{sh}}(=2\theta_{\infty})$ as a function of $b$, when Sgr A* is modeled as REC spacetimes:(a) black hole (left) (b) no-horizon spacetime (right). The dark black line corresponds to calculated values of $2\theta_{\infty}$ as a function of $b$.  The white/red  regions represent areas that are 1 $\sigma$ consistent/inconsistent with the 2017 EHT observations, indicating that the latter imposed constraints on parameter $b$.}\label{SgrAparameter}
\end{figure*}
We use the EHT observation results of M87* and Sgr A* black hole shadows to constrain the deviation parameter $b$ associated with REC black hole spacetime.  By considering the apparent radius of the photon sphere $\theta_{\infty}$ as the angular size of the black hole shadow, we constrain the deviation parameter within the 1 $\sigma$ level.  The EHT observational results are compatible with the reality of an event horizon, proving that it ruled out no-horizon spacetimes or naked singularities.
However, we still model the M87* and Sgr A* as the REC spacetimes (both black holes and no horizon) and use the EHT observations results to test the viability of these REC spacetimes. 

\paragraph{Constraints from  M87*:} We find that  the Schwarzschild black hole  ($b = 0$) casts the largest shadow with its angular diameter $\theta_{\text{sh}}=2\theta_\infty= 39.56069~\mu$as, which falls within the 1 $\sigma$ region for the black hole with mass  $M=(6.5 \pm 0.7) \times 10^9 M_\odot$ and distance of $D_{\text{OL}}=16.8$ Mpc \citep{EventHorizonTelescope:2019dse,EventHorizonTelescope:2019pgp,EventHorizonTelescope:2019ggy}. Figure \ref{M87parameter} depicts the angular diameter $\theta_{\text{sh}}$ as a function of $b$, with the black solid line corresponding to $\theta_{\text{sh}}=39~\mu$as for the REC black hole spacetime as M87*. The REC black hole spacetime metric when investigated with the EHT results of M87*  within the 1 $\sigma$ bound, constrains parameter $b$, viz., $0< b \le 0.0165174$ . Thus, based on Figure \ref{M87parameter}, within only a tiny part of parameter space, REC black hole spacetime could be a candidate for the astrophysical black holes. However, according to our results, the no-horizon spacetime does not agree with the EHT results of M87* (Figure \ref{M87parameter}). 

\paragraph{Constraints from  Sgr A*:}The EHT observation used three independent algorithms  to find out that the averaged measured value of the shadow angular diameter, which lies in the range of $\theta_{\text{sh}} \in (46.9, 50)~\mu$as, and the 1 $\sigma$ interval is $\in$ $(41.7 ,55.6)~\mu$as \citep{EventHorizonTelescope:2022xqj}. The angular diameter  $\theta_{\text{sh}} \in (41.7 ,55.6) \mu$as, which falls within the 1 $\sigma$ confidence region with the observed angular diameter of the EHT observation of Sgr A* black hole, strongly constrains the parameter $0.0 \le b \le 0.1881459$ for the REC black hole spacetime.  Thus, within the finite parameter space, the REC black hole spacetime  agrees with the EHT results of Sgr A* black hole shadow (see Figure \ref{SgrAparameter}).  However, the REC no-horizon spacetimes  are again inconsistent with the EHT observation of Sgr A*.  

\section{Discussion and Conclusion}\label{sec8}
General relativity's no-hair theorem states that the Kerr-Newman metric \citep{Newman:1965my} is the only stationary, axially symmetric, and asymptotically flat electro-vacuum solution of the Einstein equations \citep{Israel:1967wq,Israel:1967za,Carter:1971zc,Hawking:1971vc,Robinson:1975bv}. Three parameters describe it: mass, angular momentum, and electric charge.
However, the third black hole parameter in the black hole, an electric charge, is often overlooked and implicitly set identically to zero. However, both classical and relativistic processes can lead to a small nonzero charge of black holes. Zero charges are a good approximation when dealing with neutral particles and photons. Even a small charge can significantly influence the motion of charged particles in the environs of black holes \citep{Kumar:2020yem,Kumar:2020ltt,Kumar:2020sag,Zhao:2016kft,Sotani:2015ewa}. Therefore,  one should not dump electric charge a priori in an astrophysical investigation.  The gravitational effect of spacetime curvature by mass currents is the rotation of the plane of polarization for linearly polarized light rays, known as the Rytov effect \citep{rytov1938transition}. Such a gravitational rotation of the polarization plane in stationary spacetime is a gravitational analog of the electromagnetic Faraday effect \citep{Piran:1985dk,Ishihara:1988,Nouri:1999}.

Hence, we investigated the feasibility of comparing black holes from no-horizon spacetimes via strong gravitational lensing and analyzed the astrophysical consequences for several supermassive black holes considering electric charge. We look at a class of REC metrics, which are no-horizon spacetimes for parameters $b>b_E  \approx  0.226$ and can also provide spherically symmetric black hole solutions when $0<b\leq b_E$. Interestingly, we found that a photon sphere does exist for $0 < b \leq    b_P \approx 0.247$ for REC spacetimes. On the other hand, the anti-photon sphere with stable circular orbits forms for no-horizon spacetime for $b_E < b \leq b_P $. We found that the radius of the photon sphere $x_{\text{ps}}$ is a decreasing function of the $b$, while the opposite behavior is shown by the anti-photon sphere radius $x_{aps}$ and they merge at $x \approx b_E$. The deflection angle $\alpha_D$, for REC spacetime diverges at the critical impact parameter $u=u_{\text{ps}}$ and is a decreasing function of the impact parameter $u$ as well as of  the parameter $b$ for REC black holes. Interestingly, for REC no-horizon spacetime in the limit $u \to u^+_{\text{ps}}$, the deflection angle $\alpha_D$, increases rapidly in comparison when $u \to u^-_{\text{ps}}$ (cf. Figure \ref{fig12}). We  modeled compact objects Sgr A*, M87* NGC 4649 and NGC 1332 as the REC spacetimes to  evaluate the  lensing observables, viz., angular positions of asymptotic relativistic images $\theta_{\infty}$, angular separations $s$, and relative magnifications $r_{\text{mag}}$ of relativistic images. We found that $\theta_{\infty}$ decreases with the parameter $b$ for REC black holes as well as the for the no-horizon spacetime. $\theta_{\infty}$ $\in~(19.0178,26.3299)~\mu$as for Sgr A* and $\in~(14.2884,19.782)~\mu$as for M87* when the compact objects are considered as REC black holes, while $\theta_{\infty}$  $\in~(17.4462, 19.0178)~\mu$as for Sgr A* and $\in~(13.1076,14.2884)~\mu$as for M87* when they are considered as REC no-horizon spacetimes. We have taken into account the three additional images apart from $\theta_{1}$, at angular positions  $\theta_{-1}$, $\theta_{-2}$, and $\theta_{-3}$ for no-horizon spacetime. Further, we calculated the separation $s$ for black hole spacetime and $s_1, s_{-1}, s_{-2}$ and $s_{-3}$ for REC no-horizon spacetime.  The separation  of the first relativistic image from the other packed images at $\theta_{\infty}$ increases with $b$ for $0<b \le b_{E}$. It further increases attaining a peak at $b\approx0.224$ and thereafter decreases for $0.244 < b \le b_{P}$. Contrarily, $s_{-1}, s_{-2}$ and $s_{-3}$ show a decreasing behavior with the parameter $b$. The separation $s$ is much larger than that of Schwarzschild black hole for both Sgr A* and M87*, and is respectively, in the range 0.0329517-0.199342 and 0.0247571-0.149769~$\mu$as. On the other hand, $s_1 \in$ (0.1882,0.319045)~$\mu$as,  $s_{-1} \in$ (1.12743,7.8168)~$\mu$as,  $s_{-2} \in$ (0.512824, 2.04872)~$\mu$as, and $s_{-3} \in$ (0.226226,0.322697)~$\mu$as for Sgr A*  and $s_1 \in$ (0.141397,0.239703)~$\mu$as,  $s_{-1} \in$ (0.847053,5.87288)~$\mu$as,  $s_{-2} \in$ (0.385292,1.53924)~$\mu$as, and $s_{-3} \in$ (0.169967,0.242447)~$\mu$as for M87*.  Consequently, the images formed from  inside of the photon sphere are sufficiently separated from each other and some are marginally within the current resolution capability. Moreover, the relative magnification of the outermost relativistic image caused by REC black hole spacetime decreases with   $b$ such that it varies between $4.56657$ and $6.82188$ orders of magnitude.  On the contrary, $\mathcal{R}_{1} \in (0.42613,2.0822), \mathcal{R}_{-1} \in (1.42485,3.90258) $, $\mathcal{R}_{-2} \in (0.669416,3.54039) $, $\mathcal{R}_{-3} \in (-0.17369,1.79488) $.
Thus, the observables of the images formed from inside the photon sphere are comparably different, showing that the strong gravitational lensing signature of no-horizon spacetime varies qualitatively from black holes.

We have also calculated the time delay $\Delta T^s_{2,1}$ between first- and second-order relativistic primary images for 22 supermassive black holes. Considering them as REC black hole spacetime, we found,  e.g., $\Delta T^s_{2,1}$ for  Sgr A*, M87*, NGC 4649, and NGC 1332, respectively, can reach $\sim 8.8809$, $\sim 12701.8$, $\sim 9748.35$, and $\sim 3036.03$ minutes with the deviation from the Schwarzschild black hole of same mass and distance, respectively,  being $\sim 2.6159$, $\sim 4677$, $\sim 2871.35$, and $\sim 894.26$ minutes. Except for Sgr A*, the time delay can be in the hundreds of hours, which is adequate for astronomical measurements if we have enough angular resolution between two relativistic images.  

On using the  EHT results of the M87* and Sgr A* black hole shadows put a bound on the parameter $b$ for the REC spacetime, which can be estimated based on the supposition that the relativistic image is the apparent size of the shadow at $\theta_{\infty}$. When investigated with the EHT results of M87*  within the 1-$\sigma$ bound, the REC black hole spacetime metric puts a tight bound on the parameter $b$, viz., $0< b \le 0.0165174$. On the other hand, the results of Sgr A* restrict $b$ such that  $0.0 \le b \le 0.1881459$. Meanwhile, the REC no-horizon spacetimes are inconsistent with EHT observations of M87* and Sgr A*. Thus, EHT bounds on $\theta_{\text{sh}}$ of Sgr A* and M87*, within the $1 \sigma$ region, which place bounds on the parameter $b$, and hence we conclude that although the REC black holes agree with the EHT results in finite parameter space,   the corresponding REC no-horizon spacetimes are complete ruled out. The spacetime under consideration is nonrotating; this work predicts strong deflection gravitational lensing signals for REC spacetime and hints at its no-horizon spacetimes.

Finally, due to the complicated REC spacetime metric, in the present analysis, we have limited our study to the spherically symmetric case, i.e.,  overlooked spin. It is because the silhouette of the shadow, as watched at infinity,  has a shape that weakly depends on the spin of the black hole \citep{Bardeen:1973tla, EventHorizonTelescope:2020qrl}. However,  we can anticipate our results on lensing by supermassive black holes Sgr A* and M87* are valid,  and the EHT observation can also be assumed to test these spherical black holes. Meanwhile, a thorough analysis of the rotating counterpart will be a promising avenue for the future.
\section{Acknowledgments} 
 J.K. would like to thank CSIR for providing SRF.
 S.U.I and S.G.G.  would like to  thank SERB-DST for the project No. CRG/2021/005771. 
 
\bibliography{REC}
\bibliographystyle{aasjournal}
\end{document}